\documentclass[pra,twocolumn,showkeys,showpacs]{revtex4} 
\usepackage{amsmath,bm}
\usepackage{amssymb}
\usepackage{graphicx}
\usepackage{amsmath}
\usepackage{amsfonts}

\newcommand{\A}{\mathbf{A}}
\newcommand{\B}{\mathbf{B}}
\renewcommand{\b}{\mathbf{b}}

\newcommand{\E}{\mathbf{E}}
\newcommand{\D}{\mathbf{D}}

\newcommand{\e}{\mathbf{e}}
\renewcommand{\i}{\mathbf{i}}
\newcommand{\p}{\mathbf{p}}
\renewcommand{\r}{\mathbf{r}}
\newcommand{\s}{\mathbf{s}}
\renewcommand{\u}{\mathbf{u}}
\renewcommand{\v}{\mathbf{v}}
\newcommand{\z}{\mathbf{z}}
\newcommand{\x}{\mathbf{x}}
\renewcommand{\x}{\mathbf{x}}

\newcommand{\tpsi}{\widetilde\psi}

\newcommand{\tF}{\widetilde F}

\newcommand{\tG}{\widetilde G}
\newcommand{\tR}{\widetilde R}

\newcommand{\tV}{\widetilde V}

\newcommand{\bsig}{\bm{\sigma}}

\newcommand{\half}{{\textstyle \frac{1}{2}}}
\newcommand{\smallhalf}{{\scriptstyle\frac{1}{2}}}
\newcommand{\bdot}{\bm{\cdot}}

\begin{document}

\title{Zitterbewegung structure in electrons and photons }

\author{David Hestenes}
\affiliation{Department of Physics, Arizona State University, Tempe, Arizona 85287-1504}
\email{hestenes@asu.edu}
\homepage{http://geocalc.clas.asu.edu/}

\begin{abstract}

The Dirac equation is reinterpreted as a constitutive equation for singularities in the electromagnetic vacuum, with the electron as a point singularity on a lightlike  toroidal vortex. The diameter of the vortex is a Compton wavelength and its thickness is given by the electron's anomalous magnetic moment.
The photon is modeled as an electron-positron pair trapped in  a vortex with energy proportional to the photon frequency. 
The possibility that all elementary particles are composed of similar vortices is discussed.

\end{abstract}

\pacs{10,03.65.-w}
\keywords{pilot waves, Hopf fibrations, Hall effect, zitterbewegung, spacetime algebra, QED}

\maketitle

\section{Introduction}

The spectacular success of quantum electrodynamics (QED) gives physicists great confidence in Maxwell's equation on the one hand and Dirac's equation on the other, yet something is missing in relations between them. With his usual penetrating insight, Einstein focused on the crux of the problem \cite{Jaynes96,Mead97}: ``It is a delusion to think of electrons and the fields as two physically different, independent entities. Since neither can exist without the other, there is only \textit{one} reality to be described, which happens to have two different aspects; and the theory ought to recognize this from the start instead of doing things twice."  

This paper proposes a synthesis of Maxwell and Dirac theories based on a new model for singularities in the electromagnetic vacuum.  
The model is suggested by a remarkable relation between electron mass and vacuum polarization proposed by Seymour Blinder. 
The only requirement is consistency with  Maxwell's equation.
No changes in the form of Dirac or Maxwell equations are necessary, but the two are fused at the source. 
The solutions seamlessly integrates electron field and particle properties along lines proposed by de Broglie.  They answer Einstein's call for a unified electron theory with a unified  \textit{Maxwell-Dirac} theory.

Singular toroidal solutions of the Dirac equation constitute a new class of wave functions, fairly called \emph{ontic states} (or ``states of reality" as Einstein might have put it), because they have a definite physical interpretation in terms of local observables of the electron and associated deformation of the vacuum. No probabilities are involved.
Electron states are thus characterized by a literal \emph{field-particle duality}: field and particle coexist as a real physical entity. 
This appears to finesse the notorious self-energy problem. It implies there is no such thing as the electron's own field acting on itself, because particle and field are two different aspects of one and the~same~thing. 

Section II reviews the formulation of classical electromagnetic theory in terms of Spacetime Algebra (STA)  to provide a context for two important new developments. The first is Blinder's  concept of a classical vacuum singularity.
The second is Antonio Ra\~nada's discovery of toroidal solutions to Maxwell's equation. The two provide complementary inputs to a new theory of the electron and the electromagnetic vacuum.

Section III begins with a review and extension of the \textit{zitter particle model} for the electron clock in \cite{Hest90}
and updated in the preceding paper \cite{Hest19a}. Then the model is grounded with fundamental constants. That leads to Oliver Consa's  calculation  \cite{Consa18} for the electron anomalous magnetic moment and for its physical explanation. 

The stage is set in Section IV for the main subject of this paper, namely, reconstruction of the Dirac equation  as a constitutive equation for the vacuum with the electron as a point singularity.
The singularity generates the electron's Coulomb field with toroidal zitterbewegung and   magnetic field with the spin vector as its axis. 
I call this  \textit{Maxwell-Dirac} theory. It is complementary to the conventional \textit{Born-Dirac} theory
discussed in \cite{Hest19a}
in the sense that motion is described by the same Dirac equation in both.
In Born-Dirac the electron charge is inert and responds only to action of external fields. In Maxwell-Dirac the charge is active and generates an electromagnetic field. In this sense, Maxwell-Dirac may be regarded as an alternative to second quantization, though we do not directly consider  its relation to standard QED.

Section V proposes a \textit{new model for the photon} as an electron-positron pair bound in a toroidal ring with the diameter of a reduced Compton wavelength.   
The ring has quantized internal states that have been observed in photon diffraction \cite{Helmerson16}.\,That invites reconsidering in Section VI the possibility that  all elementary particles can be constructed from leptons, as proposed in a seminal analysis by Asim Barut \cite{Barut80,Barut86}.

Section VII discusses many particle systems and diffraction. Considering the enormous scope of Maxwell-Dirac theory, our treatment has many loose ends and is best regarded as a guide for further research, though the model of electron as a point particle with an inherent periodicity is an essential feature in all variants of the theory.

\section{Classical Electromagnetics}\label{sec:II}

To place our analysis in the most general context, we begin with an STA formulation of Maxwell's electrodymamics in accord with the authoritative presentation by Sommerfeld \cite{Sommerfeld52}.

In STA an\textit{ electromagnetic field} is represented by a
bivector-valued function $F = F(x)$ on spacetime, appropriately called the {\it Faraday}. 
Its split into electric and magnetic fields is given by  
\begin{equation}
 F = \mathbf{E}+i\mathbf{B}.\label{2.36}
\end{equation}
In a polarizable medium, the \textit{electromagnetic field density} is a bivector field $G = G(x)$ with the split into electric and magnetic densities given by 
\begin{equation}
 G = \mathbf{D}+i\mathbf{H}.\label{2.37}
\end{equation}
The important distinction between ``field" and ``field density" or ``excitation" is emphasized by Sommerfeld. We are interested in $G$ for describing properties of the vacuum.

The most general possible version of  {\it Maxwell's equation} for the electromagnetic field is
\begin{equation}
 \square F = J_{e}+ iJ_{m}\,,\label{2.39}
\end{equation}
where $ J_{e} = J_{e}(x)$ is the {\it electric charge current }and $ J_{m} = J_{m}(x)$ is a {\it magnetic charge current}.
 Separating vector and pseudovector parts, we get
\begin{equation}
\square\bdot F= J_{e}\label{2.40}
\end{equation}
and
\begin{equation}
\square\wedge F= iJ_{m}.\label{2.41}
\end{equation}
Using the \textit{duality} between divergence and curl 
\begin{equation}
(\square\wedge F)i=\square\bdot (iF)
\label{2.42}
\end{equation}
and the anticommutivity of the pseudoscalar with vectors,
the latter equation can be written
\begin{equation}
\square\bdot (iF) =J_{m}.\label{2.43}
\end{equation}

In the most general case, the Faraday
 $ F $ can be derived from a ``complex'' pair of vector potentials $ A=A(x) $ and $ C=C(x) $, so we have 
\begin{equation}
 F = \square (A+Ci). \label{2.43a}
\end{equation}
The scalar and pseudoscalar parts of this equation give us
\begin{equation}
\square \bdot A =0=\square \bdot C, \label{2.43b}
\end{equation}
while equations (\ref{2.40}) and  (\ref{2.43}) give us separate equations for fields produced by electric and magnetic charges:
\begin{equation}
\square^2 A= J_{e},\label{2.43c}
\end{equation}
\begin{equation}
\square^2 C= J_{m}.\label{2.43d}
\end{equation}
Though we shall dismiss the magnetic monopole current $J_{m}$ as unphysical, we shall see good reason to keep the complex vector potential for radiation fields.


Squaring the Faraday gives us scalar and pseudoscalar invariants which can be expressed in terms of electric and magnetic fields:
\begin{equation}
 F^2 =(\E+i\B)^2 =\E^2 -\B^2+2i\E\bdot\B. \label{2.43f}
\end{equation}
As first shown in \cite{Hest66}, if either of these invariants is nonzero, they can be used to put the Faraday in the unique invariant form:
\begin{equation}
 F = \mathbf{f} e^{i\varphi}=\mathbf{f}\cos\varphi +i\mathbf{f}\sin\varphi ,  \label{2.43g}
\end{equation}
where the exponential specifies a duality transformation through an angle given by
\begin{equation}
 \tan{2\varphi}=\frac{2\E\cdot\B}{\E^2 -\B^2}=\frac{
 iF\wedge F}{F\bdot F}.  \label{2.43h}
\end{equation}
Note that (\ref{2.43g}) determines a rest frame in which the electric and magnetic fields are parallel without using a Lorentz transformation. In addition, the squared magnitude of $\mathbf{f}$ is 
\begin{equation}
\mathbf{f}^2=[(\E^2+\B^2)^2-4(\E\boldsymbol{\times}\B)^2]^\half,  \label{2.43i}
\end{equation}
which is an invariant of the Poynting vector for $F$.

A null field can also be put in the form 
\begin{equation}
 F = \mathbf{f}e^{i\varphi}\quad\hbox{with}\quad \mathbf{f}^2=0,  \label{2.43j}
\end{equation}
so we can write
\begin{equation}
\mathbf{f}= \e +i\b=\e(1+\hat{\mathbf{k}}) =\mathbf{f}\hat{\mathbf{k}}. \label{2.43k}
\end{equation}
Hence we can put the null field  into the form
\begin{equation}
F = \mathbf{f}e^{i\varphi}
=\mathbf{f}e^{i\hat{\mathbf{k}}\varphi},  \label{2.43l}
\end{equation}
showing that the duality rotation is equivalent to a rotation of vectors in the null plane. Of course, this is significant for  description of radiation fields. 

\subsection{Conservation Laws}

Using the identity 
\begin{equation}
\square\bdot(\square\bdot F)=(\square\wedge\square)\bdot F=0, \label{2.45}
\end{equation}
from (\ref{2.40}) and (\ref{2.43}), we get the current conservation laws
\begin{equation}
 \square\bdot J_{e}=0\quad\hbox{and}\quad  \square\bdot J_{m}=0.\label{2.46}
\end{equation}

An energy-momentum tensor $T(n)=T(n(x),x)$ describes the energy-momentum flux in direction of a unit normal $n$ at spacetime point $x$.  
As discussed elsewhere \cite{Hest03b,Lasenby93}, the electromagnetic energy-momentum tensor 
is given by
\begin{equation}
T(n) = \frac{1}{2}\langle Fn\tG\rangle_{1}=\frac{1}{4}[Fn\tG+Gn\tF].
\label{2.47}
\end{equation}
where $ n $ is a unit normal specifying the direction of flux.
With some algebra, the tensor can be expressed in the alternative form, which separates normal and tangential fluxes:
\begin{equation}
T(n) = \frac{1}{2}[(G\bdot\tF)n+G\bdot(n\bdot\tF)+(G\bdot n)\bdot\tF]. 
\label{2.48}
\end{equation}
Whence,
\begin{equation}
\partial_{n}T(n) = \partial_{n}\bdot T(n)+\partial_{n}\wedge T(n)=0. \label{2.49}
\end{equation}
 The vanishing of both scalar and bivector parts in this expression tells us the linear function $ T(n) $ is traceless and symmetric.
 
The divergence of the energymomentum tensor is given by
\begin{equation}
\dot{T}(\dot{\square}) = \half\langle \dot{F}\dot{\square}\tG+F\square \tG\rangle_{1}=\half[  G\bdot J_{e}+F\bdot J_{f}], \label{2.50}
\end{equation}
where current $J_{f}$ includes any polarization or magnetization currents.
Note use of the overdot to indicate differentiation to the left.

Physical interpretation of the energymomentum tensor is perhaps facilitated by using a $ v $-\textit{split} to put it in the form     
\begin{equation}
T(v) =\frac{1}{4}[FG^{\dagger}+GF^{\dagger}]v. 
\label{2.51}
\end{equation}
From this we find the energy density
\begin{equation}
T(v)\bdot v =\half F\bdot\tG=\half (\E\bdot\D+\B\bdot \mathbf{H}),  \label{2.52}
\end{equation}
and the momentum density
\begin{equation}
T(v)\wedge v =\half(\D\boldsymbol{\times} \B +\E\boldsymbol{\times} \mathbf{H}).  \label{2.53}
\end{equation}
For a unit normal $ \mathbf{n} =nv $ orthogonal to $v$, we have
\begin{equation}
T(n)v =\frac{1}{4}[F\mathbf{n}G^{\dagger}
+G\mathbf{n}F^{\dagger}],  \label{2.54}
\end{equation}
so
\begin{equation}
T(n)\bdot v =\half(\D\boldsymbol{\times} \B +\E\boldsymbol{\times} \mathbf{H})\bdot\mathbf{n},  \label{2.55}
\end{equation}
and the spatial flux in direction $\mathbf{n}  $ is
\begin{equation}
\bm{T}(\bm{n} ) = T(n)\wedge v 
=\frac{1}{4}\langle\E\mathbf{n}\D+\B\mathbf{n}\mathbf{H}
\rangle_{\bm{1}}. \label{2.56}
\end{equation}
The treatment of energymomentum conservation in this subsection is completely general, applying to models of the electromagnetic vacuum considered next as well as material media. It sets the stage for specific applications considered next as well as extensions to be considered in the future.

\subsection{The Classical Vacuum}

In a thorough analysis of constitutive equations in Maxwell's electrodynamics, E. J. Post \cite{Post72} identified a hitherto unrecognized degree of freedom in Maxwell's equation for the vacuum. Regarding the vacuum as a dielectric medium with variable permittivity $\varepsilon = \varepsilon(x)$ and permeability $\mu = \mu(x)$ at each spacetime point $x$, Maxwell's condition for the propagation of light in a vacuum is given by
\begin{equation}
\varepsilon\mu=1/c^{2}=\varepsilon_{0}\mu_{0}.\label{3.1}
\end{equation}
Obviously, this leaves the \emph{impedance}
\begin{equation}
Z=\sqrt{\frac{\mu}{\varepsilon}}=Z(x)\label{3.2}
\end{equation}
as an undetermined function. To ascertain its value, Post further argues that charge should be regarded as an independent unit $q_{e}$ rather than the derived unit $e$.
The standard rule for changing units is
\begin{equation}
e^{2}=\frac{q_{e}^{2}}{4\pi\varepsilon_{0}}.\label{3.3}
\end{equation}
Accordingly, the \emph{fine structure constant} is given by
\begin{equation}
\alpha_{e}=\frac{e^{2}}{\hbar c}=\frac{q_{e}^{2}}{4\pi\varepsilon_{0}\hbar c}=\frac{q_{e}^{2}}{4\pi\hbar}\sqrt{\frac{\mu_{0}}{\varepsilon_{0}}}.\label{3.4}
\end{equation}
Hence, as Post observes, the fine structure constant can be expressed as a ratio of two generally invariant impedances:
\begin{equation}
\alpha_{e}=\frac{Z_{0}}{Z_{H}},\quad\hbox{where}\quad
Z_{0}=\sqrt{\frac{\mu_{0}}{\varepsilon_{0}}},\quad
 Z_{H}=\frac{2h}{q_{e}^{2}}. \label{3.5}\end{equation}
This suggested to Post
 that the \emph{Hall impedance }$Z_{H}$ is an intrinsic property of the electromagnetic vacuum.

Blinder \cite{Blinder01a,Blinder03} has shown that polarization of the vacuum in the neighborhood of a classical electron is uniquely determined by the very simple assumptions that (1) the energy density of the electron field is proportional to the charge density, and (2) the total energy in the field determines the electron mass. We review Blinder's argument to serve as a guide for generalizing it to the Dirac electron.

For a point charge in its rest frame the electric field $\E$ and electric displacement $\D$ are given by
\begin{equation}
\E=\frac{q_{e}}{4\pi\varepsilon(r)}\frac{\r}{r^{3}}, \qquad
\D=\varepsilon(r)\E=\frac{q_{e}}{4\pi}\frac{\r}{r^{3}},\label{3.6}
\end{equation}
where $r=|\r|$. The total energy in the field is
\begin{equation}
\hspace*{-.2in}W=\frac{1}{2}\int \E\bdot\D\, d^{3}r=\frac{1}{32\pi^{2}}\int_{0}^{\infty}\frac{1}{\varepsilon(r)}
\frac{q_{e}^{2}}{r^{4}}\,4\pi r^{2}dr.\label{3.7}
\end{equation}
The charge density $\varrho$ is determined by
\begin{equation}
\frac{\varrho}{\varepsilon_{0}}=\boldsymbol{\nabla}\bdot \E=\frac{-q_{e}}{4\pi}\left[\frac{\varepsilon'(r)}{r^{2}[\varepsilon(r)]^{2}}
+\frac{\delta^{3}(\r)}{\varepsilon(0)}\right]\label{3.8}
\end{equation}
with $\varepsilon'=\partial_r\, \varepsilon$  and the normalization
\begin{equation}
\int_{0}^{\infty}\varrho(r)\,4\pi r^{2}dr
=-q_{e}\int_{0}^{\infty}
\frac{\varepsilon_{0}\varepsilon'(r)dr}{[\varepsilon(r)]^{2}}=q_{e},\label{3.9}
\end{equation}
which require that $\varepsilon(\infty)=\varepsilon_{0}$ and $\varepsilon(0)=\infty$.

Finally, assuming that $W=m_{e}c^{2}$ and the charge density in (\ref{3.8}) is proportional to the energy density in (\ref{3.7}), we get
\begin{equation}
\frac{-\varepsilon_{0}\varepsilon'(r)}{r^{2}[\varepsilon(r)]^{2}}
=\frac{q_{e}^{2}}{32\pi^{2}m_{e}c^{2}\varepsilon(r)r^{4}}.\label{3.10}
\end{equation}
Whence,
\begin{equation}
\varepsilon(r)=\varepsilon_{0}\exp{\left(\frac{\lambda_{0}}{r}\right)},\label{3.11}
\end{equation}
where
\begin{equation}
\lambda_{0}=\frac{q_{e}^{2}}{8\pi \varepsilon_{0}}\frac{1}{m_{e}c^{2}}=\frac{1}{2}\frac{e^{2}}{m_{e}c^{2}}\label{3.12}
\end{equation}
is recognized as half the \emph{classical electron radius} and puts that quantity into new perspective as a radius of vacuum polarization.

Now we have an explicit expression for the vacuum charge density:
\begin{equation}
\boldsymbol{\nabla}\bdot\E=\frac{q_{e}}{4\pi\varepsilon_{0}}\frac{\lambda_{0}}{r^{4}}\,
e^{-\lambda_{0}/r}=\frac{\varrho(r)}{\varepsilon_{0}}.\label{3.13}
\end{equation}
Sommerfeld \cite{Sommerfeld52} emphasizes that this does not have the dimensions of charge, and he interprets it simply as ``divergence of the electric field." Of course, he was not privy to our notion of vacuum polarization or its expression as a 
manifestly nonsingular quantity. However, the dimensions of charge and its singularity are explicit in
\begin{equation}
\boldsymbol{\nabla}\bdot\D=\frac{-q_{e}}{4\pi}\,\,\boldsymbol{\nabla}^{\,2}\left(\frac{1}{r}\right)
=q_{e}\delta^{3}(\r).\label{3.14}
\end{equation}
Also note that $\E = -\boldsymbol{\nabla}\varphi_{e},$ where
\begin{equation}
\varphi_{e}(r)=\frac{-q_{e}}{4\pi\varepsilon(r)\lambda_{0}}.\label{3.14a}
\end{equation}
This suggests a straightforward generalization of Blinder's argument.

By interpreting the variable $r$ as the retarded distance  between each spacetime point $x$ and the path $z=z(\tau)$ of a point charge with velocity $v=\dot{z}=c^{-1}dz/d\tau$, the simple form  for the scalar potential in (\ref{3.14a}) leads immediately to the following generalization of the Li\`{e}nard-Wiechert potential:
\begin{equation}
A(x)=\frac{q_{e}}{4\pi\varepsilon(r)}\frac{v}{\lambda_{0}}.\label{3.15}
\end{equation}
Let's call this the \textit{``Coulomb vector potential''} to emphasize its relation to the Coulomb scalar potential.
The retarded distance is defined explicitly by
\begin{equation}
r=(x-z(\tau))\bdot v =|\r|\quad\hbox{with}\quad
\r=(x-z(\tau))\wedge v\label{3.16}
\end{equation}
subject to the constraint
\begin{equation}
(x-z(\tau))^{2}=0.\label{3.17}
\end{equation}
Hence, it generates the electromagnetic field
\begin{equation}
F=\square\wedge A=\frac{q_{e}}{4\pi\varepsilon(r)}\left\{
 v\wedge\square\frac{1}{r}
+\frac{\square\wedge v}{\lambda_{0}}\right\}.\label{3.19}
\end{equation}
To evaluate the derivatives, we use the constraint (\ref{3.17}), which implies proper time $\tau=\tau(x)$ as a function of position with gradient 
\begin{equation}
\square \tau=\frac{x-z}{c\,r}\equiv k,\label{3.19aa}
\end{equation}
where null vector $k$ is independent of distance $r$.
Consequently, the curl of $v=v(\tau(x))$ has the simple form
\begin{equation}
\square\wedge v=\square \tau \wedge \partial_{\tau}v =c\,k\wedge \dot{v}.\label{3.19a}
\end{equation}
Similiarly, the gradient of (\ref{3.16}) gives us
\begin{align}
\square r&=v (v\wedge \square r)+\square\tau \,(x-z)\bdot \dot{v}c  \notag \\
&=v\hat{\r}+crkk\bdot\dot{v}.\label{3.19ab}
\end{align}
Inserting all this into (\ref{3.19}), we get 
\begin{align}
F&=\square\wedge A  \notag \\ &=\frac{q_{e}}{4\pi\varepsilon(r)}\left\{\frac{\r}{r^{3}}
+\frac{ck\bdot\dot{v}}{r}v\wedge k
+\frac{ck\wedge\dot{v}}{\lambda_{0}}\right\}.\label{3.19b}
\end{align}
This differs from the classical retarded field \cite{Hest93b} only in the last term, wherein the distance $r$ in the denominator is replaced by
 $\lambda_{0}$. At first sight this seems wrong, because a spherical electromagnetic wave surely attenuates with distance.
That may be why Blinder did not associate his vacuum impedance with the electron's vector potential.
On the other hand, if Blinder's idea has any relevance to quantum mechanics it must be expressed through the vector potential, because that is the only mechanism for electromagnetic interaction.
Indeed, if the last term contributes to photon emission, it should not depend on distance.
Let us therefore withhold judgment until we examine how the vacuum impedance might fit into quantum mechanics.

To wind up our discussion of classical theory, we 
define a generalized displacement field by $G=\varepsilon F$. Whence its divergence for a charge at rest is
\begin{equation}
\hspace*{-.21in}\square \bdot G=q_{e}\delta^{3}(\r)v=q_{e}\int_{-\infty}^{{\,\infty}}
dz\delta^{4}(x-z(\tau))=J_{e},\label{3.20}
\end{equation}
in agreement with (\ref{3.14}) and (\ref{2.40}).
These expressions for $ F $ and $G=\varepsilon F$ suffice to fit Blinder's model for a point charge in the vacuum into the general formulation of classical electrodynamics in the preceding section. 

For a more realistic model of the electron, we need to incorporate electron spin and magnetic moment as well. Blinder \cite{Blinder03} tried that with a dipole model of the magnetic moment, and he deduced an exponential form for the magnetic permeability $\mu$ analogous to that for $\varepsilon$, but with a different functional dependence. However, that approach is
inconsistent with Maxwell's condition (\ref{3.1}) relating $\mu$ and $\varepsilon$ to the speed of light. Instead, from now on, we impose Maxwell's condition by introducing the dimensionless vacuum impedance

\begin{equation}
\epsilon(x)=\sqrt{\frac{\varepsilon}{\mu}}\sqrt{\frac{\mu_{0}}{\varepsilon_{0}}}=\frac{\varepsilon}{\varepsilon_{0}}=\frac{\mu_{0}}
{\mu}=e^{\lambda_{0}/r}\label{3.21}
\end{equation}
 I call the inverse of this function with some given value for $\lambda_{0}$ a \emph{Blinder function} in recognition of Blinder's seminal contribution.

\subsection{ Radiation Fields}

A remarkable new family of null solutions to Maxwell's equation was discovered by Antonio Ra\~{n}ada in 1989 \cite{Ranada89,Ranada92} and subsequently reviewed and extended in \cite{Ranada14,Ranada12}.
He called them ``knotted radiation fields'' because electric and magnetic field lines are interlocked in toroidal knots that persist as fields propagate.
As beautifully described in \cite{Irvine08,Irvine13}, this has opened up an exciting new thread of research on electromagnetic radiation.

Many researchers have discovered  advantages in formulating the radiation field as a ``complex vector'' $F=\E+i\B$ with the null condition $F^2=F\circ F=0$, but only Enk \cite{Enk13} has formulated it with geometric algebra to show that the imaginary unit must be interpreted as the unit pseudoscalar of spacetime. 
We shall demonstrate that STA has further advantages   preparing the way for extending the theory
to include toroidal fields described by the Dirac equation in a later Section.  

As emphasized by Kholodenko \cite{Kholodenko16a,Kholodenko16b},
a crucial element in an electromagnetic knot is a ``self-generating'' vector field $\v=\v(\x)$ described by the ``eigenvector equation''
\begin{equation}
\boldsymbol{\nabla\times}\v=\kappa \v.
\label{3.22}
\end{equation}
This equation has been known since the nineteenth century as the {\it Beltrami equation} and employed to model vorticity in fluids. It appears again in magnetohydrodynamics, where it is called the {\it force-free equation.} And in superconductivity it is known as the {\it London equation.}
Kholodenko \cite{Kholodenko13} has reviewed the vast literature on the subject across mathematics as well as physics from the unifying perspective of {\it contact geometry}.

The Beltrami equation is even inherent in the free field Maxwell equation 
\cite{Kholodenko16b}.
Indeed, using the spacetime split, Maxwell's equation  can be written 
\begin{equation}
\partial_0F=-\boldsymbol{\nabla}F .\label{3.23}
\end{equation}
For the unique reference frame and form specified by (\ref{2.43g}), we can write
\begin{equation}
 F(\x,t) = \v(\x,t) e^{i\varphi(\x,t)} .  \label{3.24}
\end{equation}
Inserting this into Maxwell's equation and using an overdot for the time derivative, we have 
\begin{equation}
( \dot{\v}+i\dot{\varphi}\v) e^{i\varphi}=-( \boldsymbol{\nabla} 
\v+i(\boldsymbol{\nabla}\varphi) \v) e^{i\varphi} .  \label{3.25}
\end{equation}
Cancelling exponentials and separately equating even and odd parts, we get
\begin{equation}
+i\dot{\varphi}\v =- \boldsymbol{\nabla} 
\v ,  \label{3.26}
\end{equation}
and
\begin{equation}
\dot{\v}=-i(\boldsymbol{\nabla}\varphi) \v .  \label{3.27}
\end{equation}
Writing $\kappa=-\dot{\varphi}$, we put (\ref{3.26}) in the form
\begin{equation}
+\kappa\v =-i \boldsymbol{\nabla} 
\v =\boldsymbol{\nabla} \boldsymbol{\times} \v.  \label{3.28}
\end{equation}
As advertised, this is precisely the  Beltrami equation, with a side condition $\boldsymbol{\nabla} \bdot\v=0$. If $\kappa$ is constant, we can differentiate again to get an equation of familiar ``Helmholtz type:''
\begin{equation}
 (\boldsymbol{\nabla}^2 +\kappa^2) 
\v=0 .  \label{3.29}
\end{equation}
Time dependence of the vector field is governed by eq. (\ref{3.27}), which can be written
\begin{equation}
\dot{\v}=- \v\boldsymbol{\times}\boldsymbol{\nabla}\varphi ,  \label{3.30}
\end{equation}
with the side condition $\v\bdot\boldsymbol{\nabla}\varphi =0$. Besides identifying a role for Beltrami's equation, this analysis serves to demonstrate some advantages of geometric calculus and the way it  incorporates standard vector calculus.

As emphasized by Enk \cite{Enk13}, it follows immediately from Maxwell's equation (\ref{3.23}), that there is a whole hierarchy of ``parabivector'' fields
$F_n\equiv\boldsymbol{\nabla}^n F$  that satisfy 
\begin{equation}
\partial_0F_n=-\boldsymbol{\nabla}F_n .\label{3.30a}
\end{equation}
In particular, with $ F=\boldsymbol{\nabla} F_0$
we can define a vector  potential
\begin{equation}
F_0= \A +i\mathbf{C}\label{3.30b}
\end{equation}
so that $\A$ is the usual magnetic vector potential given by 
\begin{equation}
i\B= \boldsymbol{\nabla}\A= i( \boldsymbol{\nabla}\boldsymbol{\times}\A) \label{3.30c}
\end{equation}
with $\boldsymbol{\nabla}\bdot\A=0$, and $\mathbf{C}$ is an analogous vector potential for the electric field $\E$. 

For radiation fields the relation to Beltrami's equation is especially simple. When $F^2 =0,$ we can write
\begin{equation}
F= \E +i\B = \rho (\e +i\b)e^{i\varphi}, \label{3.31}
\end{equation}
where $\rho$ is a single scale factor for both electric and magnetic fields so we can set $\e^2 = \b^2=1$.
Then, from (\ref{2.51}), the {\it Poynting paravector } $\mathcal{P}$ is given by
\begin{equation}
\mathcal{P} =\frac{1}{2}FF^{\dagger}=\rho^2[\frac{1}{2}(\e^2 +\b^2) -i\e\b] =\rho^2(1+\e\boldsymbol{\times}\b ). 
\label{3.32}
\end{equation}
For monochromatic radiation, all the time dependence is in the phase, so we can write $\kappa=-\dot{\varphi}$  as before, and Maxwell's equation
(\ref{3.23}) becomes
\begin{equation}
\boldsymbol{\nabla}F= i\kappa F .\label{3.32a}
\end{equation}
Defining  
``complex'' inner and outer products in terms of commutator and anticommutator parts,
we can split this into two equations
\begin{equation}
\boldsymbol{\nabla}\boldsymbol{\times}F= \kappa F ,\label{3.32b}
\end{equation}
\begin{equation}
\boldsymbol{\nabla}\circ F= 0 .\label{3.32c}
\end{equation}
Thus we see Beltrami's equation in a more fundamental role.
 Further, defining $F_n=\boldsymbol{\nabla}^n F$,  we get
\begin{equation}
\boldsymbol{\nabla}F_n= (i\kappa)^n F_n \label{3.32d}
\end{equation}
from (\ref{3.32a}) with constant $\kappa$.
Thus we have a whole nest of Beltrami fields.

With this prelude on the structure of radiation fields, we turn to 
Ra\~{n}ada's seminal 
insight into the toroidal structure of magnetic fields provided by the celebrated  \textit{Hopf fibration}. 

Hopf studied smooth maps from the 3-sphere $\mathcal{S}^3$ onto the 2-sphere $\mathcal{S}^2$ using classical techniques of complex variable theory. However, it is simpler and more informative to exploit the fact that
$\mathcal{S}^3$ is a 3-dimensional manifold isomorphic to the group $SU(2)=\{U\}$ of unitary quaternions or \textit{rotors}, to use a more descriptive term.  
 
Accordingly, we define a  
\textit{Hopf map} as a rotor-valued function $U =U(\r)$ defined on a dimensionless representation of physical space
$\mathcal{R}^3$ with a fixed origin $\r=0.$ 
Expressed in terms of {\it Euler variables}  $(u_0 , \u) $ defined and discussed in \cite{Hest86}, the rotor is normalized  by
\begin{equation}
U=u_0 +i\u \quad \hbox{with}\quad UU^{\dagger}=u_0^2 +\u^2 =1.\label{3.34}
\end{equation}
This determines the orientation of an orthonormal frame of vectors 
\begin{equation}
\e_k=U\bsig_{k}U^{\dagger},
\label{3.35}
\end{equation}
though the Hopf map requires only that one of them, say $\bsig_3 $ (designated by dropping the subscript),
serves as a \textit{pole} $\bsig$ and unit normal $\mathbf{n}=U\bsig U^{\dagger}$ for the sphere $\mathcal{S}^2$.

Accordingly, fixing the pole $\bsig$ on $\mathcal{S}^2$ determines a smooth mapping $\mathbf{n} (\r)$ of a unit normal on the surface specified by the rotor function $U(\r)$: 
\begin{align}
\mathbf{n} &=U\bsig U^{\dagger}= u_0^2\bsig +2u_0\bsig \boldsymbol{\times}\u + \u\bsig\u \notag\\
&= (u_0^2 - \u^2)\bsig +2[u_0\bsig \boldsymbol{\times}\u+(\u\bdot\bsig)\u], \label{3.35a}
\end{align}
where the identity $\u\bsig =-\bsig\u +2\u\bdot\bsig$ has been used to reorder noncommuting vectors. 
It is crucial to note, however, that the function $\mathbf{n}(\r) $
in (\ref{3.35a}) is uniquely specified by the rotor function $U(\r)$ only up to a rotation about the pole, as specified by
\begin{equation}
U_\varphi\bsig U_\varphi^{\dagger} =\bsig \quad \hbox{where} \quad U_\varphi=e^{i\bsig \varphi/2 }.
\label{3.35aa}
\end{equation}
Let's call this \textit{toroidal gauge invariance}.
This implies that $\mathbf{n}=UU_\varphi\bsig  U_\varphi ^{\dagger} U^{\dagger}=U\bsig  U^{\dagger},$
and thus reduces the degrees the freedom for $\mathbf{n}(\r)$ from three to two, while maintaining its smooth covering of the entire sphere.
This completes our formulation of the Hopf map using rotors in  geometric algebra. 

Ra\~nada had the great insight to use the Hopf map to model magnetic field lines. The essential idea is already contained in Hopf's original example for a map.
Hopf recognized that  $\mathcal{S}^3$ is isomorphic to $\mathcal{R}^3$  by  stereographic projection, as  expressed by 
\begin{equation}
2\x/\lambda_0=\frac{\u}{1-u_0},
\label{3.41}
\end{equation}
where $\lambda_0$ is a length scale factor.
Using  (\ref{3.34}) and writing $2\r=\x/\lambda_0$, this can be inverted to give
\begin{equation}
\u=\frac{2\r}{r^2+1},\quad u_0=\frac{r^2-1}{r^2+1}\quad \hbox{with}\quad r^2=\r^2.\label{3.42}
\end{equation}
Thus, we have the explicit function
\begin{equation}
U(\r)=(r^2+1)^{-1}[(r^2-1)+
2i\r].\label{3.43}
\end{equation} 
This can be inserted into (\ref{3.35a})
to give us an explicit example of a Hopf map, which has been thoroughly studied in 
\cite{Arrayas17}.

Actually, we can eliminate the stereographic projection to produce an even simpler version of the Hopf map where 
the rotor is normalized  by
\begin{equation}
U(\r)=\sqrt{\lambda}(1 +i\r) \quad \hbox{with}\quad UU^{\dagger}=\lambda(1 +\r^2) =1.\label{3.35ab}
\end{equation}
Then, the normal map  $\mathbf{n} (\r)$ on $\mathcal{S}^2$ is given by
\begin{align}
\mathbf{n} &=U\bsig U^{\dagger}= \lambda\{\bsig +2\bsig \boldsymbol{\times}\r + \r\bsig\r\} \notag\\
&=\lambda \{(1 - \r^2)\bsig +2[\bsig \boldsymbol{\times}\r+(\r\bdot\bsig)\r]\}. \label{3.36ab}
\end{align}
Thus, the normalization for the Hopf map is completely determined by the simple scaling factor $\lambda(\r)=1/(1+r^2)=1/(1+\r^2)$.
The significance of this remarkable scaling factor was implicit in Ra\~nada's treatment of knotted radiation fields from the beginning, as is evident in his treatment of a static magnetic field in \cite{Ranada14,Arrayas17}. In fact, Hopf's original example already suffices to model a magnetic field with a suitable power of the scale factor, as we now demonstrate for purposes of comparison.    

Consider the following candidate vector potential: 
\begin{equation}
\A(\r)=\half \lambda^2(\bsig\boldsymbol{\times}\r+\bsig).\label{3.43b}
\end{equation}
Differentiating 
\begin{equation}
\boldsymbol{\nabla} \lambda^2=-2\lambda^3\boldsymbol{\nabla}r^2=-4\lambda^3\r
\label{3.44}
\end{equation}
and
\begin{equation}
\boldsymbol{\nabla} (\bsig\boldsymbol{\times}
\r)=i\boldsymbol{\nabla} (\bsig\boldsymbol{\wedge}
\r)=2i\bsig,\label{3.45}
\end{equation}
we obtain
\begin{equation}
\boldsymbol{\nabla} \A=-2 \lambda^3\r(\bsig\boldsymbol{\times}\r+\bsig)
+\lambda^2i\bsig.\label{3.46}
\end{equation}
Whence
\begin{equation}
\boldsymbol{\nabla} \A=-2 \lambda^3i[\r^2\bsig-\r(\r\bdot\bsig)+i\r\bsig]
+\lambda^2i\bsig.\label{3.47}
\end{equation}
The scalar part of this equation gives us a single term
\begin{equation}
\boldsymbol{\nabla}\bdot \A=2 \lambda^3\r\bdot\bsig,
\label{3.48}
 \end{equation}
which we will need to eliminate to achieve toroidal gauge invariance.
For the moment, though, we are only interested in the curl of the vector potential $\boldsymbol{\nabla}\boldsymbol{\times} A =i\boldsymbol{\nabla}\boldsymbol{\wedge} A$.
Accordingly, from (\ref{3.47}) we get 
\begin{align}
\boldsymbol{\nabla}\boldsymbol{\times} \A &=-2 \lambda^3[\r^2\bsig-\r(\r\bdot\bsig)+\bsig\boldsymbol{\times}\r]+\lambda^2\bsig\notag\\
&= \lambda^3[2\bsig\boldsymbol{\times}\r+2\r(\r\bdot\bsig)-2r^2\bsig+(1+r^2)\bsig]
\notag\\
&=\lambda^3[2\bsig\boldsymbol{\times}\r+2\r(\r\bdot\bsig)+(1-r^2)\bsig].\label{3.50}
\end{align}
This is proportional to the normal field $\mathbf{n}(\r)$ 
defined in (\ref{3.36ab}).
Hence, with a suitable choice of units, we can identify it with a magnetic field
\begin{equation}
\B=\boldsymbol{\nabla}\boldsymbol{\times} \A=\lambda^2\mathbf{n}(\r) =\lambda^2U\bsig  U^{\dagger}.  
\label{3.51a}
\end{equation}

To make the vector potential
(\ref{3.43b}) gauge invariant, we simply need to eliminate terms orthogonal to the pole $\bsig.$ This is achieved by  
\begin{equation}
\A(\r)=\half \lambda^2(\bsig+\bsig\boldsymbol{\times}(\bsig\boldsymbol{\times}\r)),\label{3.52}
\end{equation}
where the last term is just a fancy way of writing
\begin{equation}
\bsig\boldsymbol{\times}(\bsig\boldsymbol{\times}\r)=\bsig\bdot(\bsig\boldsymbol{\times}\bsig) =\r-(\bsig\bdot\r)\bsig\equiv\r_{\perp}.\label{3.53}
\end{equation}
This now satisfies the condition (\ref{3.35aa}) for toroidal gauge invariance:
\begin{equation}
U_\varphi\A(\r)U_\varphi^{\dagger}=\A(\r).\label{3.54}
\end{equation}
And the toroidal magnetic field is given still by
\begin{equation}
i\B=\boldsymbol{\nabla}\boldsymbol{\wedge} \A =\lambda^2Ui\bsig  U^{\dagger}.  
\label{3.55}
\end{equation}
This prepares us for a straightforward generalization to toroidal radiation fields. 

As Ra\~nada has  demonstrated \cite{Ranada92}, the  structure in a monochromatic radiation field can be generated by an orthogonal pair of static vector potentials, as specified by  (\ref{3.30b}),
 where
\begin{equation}
F_0^2=(\A +i\mathbf{C})^2=2i\A\mathbf{C}=2\mathbf{C}\boldsymbol{\times}\A.\label{3.56}
\end{equation} 
Whence the radiation field has the form
\begin{equation}
F= (\boldsymbol{\nabla}F_0) e^{i\varphi}. \label{3.57}
\end{equation} 
Despite its appearance, this quantity is a relativistic invariant.
Our main interest in this result is its relevance to
modeling the photon, which is considered in a subsequent Section.

\label{3.32e}
\label{3.32f}
\label{3.2h}

.

\subsection{Geometric Calculus and differential forms}

Cartan's calculus of differential forms is used widely in mathematics and increasingly in physics, despite some significant drawbacks. Consequently, it is worth pointing out here that there is a more general  {\it Geometric Calculus} (GC) that articulates smoothly with standard vector calculus and applies equally well to spinor-valued functions. As a detailed exposition of GC is given in \cite{Hest93b,Hest84}, it suffices here to illustrate  how it relates
to differential forms in the simplest case of applications to electrodynamics.

In GC, the concept of \textit{directed integral} is fundamental, and the volume element for a $k$-dimensional integral is a (simple) $k$-vector $d^k\r=I_kd^kr$ with magnitude $|d^k\r|=d^kr$ and direction at $\r$ given by unit $k$-vector $I_k=I_k(\r )$. For a closed $k$-dimensional surface $\mathcal{S}^k$ with boundary $\mathcal{B}^k$, The {\it Fundamental Theorem of Integral Calculus} has the general form
\begin{equation}
\int_{\mathcal{S}^k}d^{k}\r'
\bdot\boldsymbol{\nabla}' f(\r'-\r)
=\oint_{\mathcal{B}^k} d^{k-1}\r' f(\r'-\r),\label{3.70}
\end{equation}
where $f(\r'-\r)$ is an arbitrary (differentiable)  function, not necessarily scalar-valued. This reduces to the fundamental theorem for differential forms when the integrands are scalar-valued.  Note that the inner product with the volume element projects away any component of the vector derivative normal to the surface.

For a multivector-valued function $A=A(\r)$ with $k$-vector parts
$A_k=<A(\r)>_k$, 
a \textit{differential $k$-form}, or just a ``\textit{$k$-form},'' can be defined by 
\begin{equation}
\alpha_k = <d^{k}\r A(\r)>=
d^{k}\r\bdot A_k.\label{3.71}
\end{equation}
The \textit{exterior derivative} of a $k$-form is a $(k-1)$-form defined by 
\begin{equation}
d\alpha_k = <d^{k}\r\bdot\boldsymbol{\nabla} A>=
(d^{k}\r\bdot\boldsymbol{\nabla})\bdot A_{k-1}.\label{3.72}
\end{equation}
The \textit{Hodge dual} can be defined (up to a choice of sign for the pseudoscalar $i$) by
\begin{equation}
*\alpha_k = <d^{k}\r A(\r)i>=
d^{k}\r\bdot A_{n-k}.\label{3.73}
\end{equation}
Here of a couple of important examples of differential forms.

The unit ``outward'' normal $\mathbf{n}$ of the directed area element $d^{2}\r$ is defined by $d^{2}\r=-i\mathbf{n}\,d^{2}r$.
Accordingly, the element of flux for a magnetic field $\B(\r)$ is given by 
\begin{equation}
 d^{2}\r\bdot (i\B)= \B\bdot \mathbf{n}  \,d^{2}r
 =d^{2}\r\bdot\boldsymbol({\nabla}
 \boldsymbol{\wedge}\A) ,\label{3.74}
\end{equation}
which integrates to a familiar form of Stokes' theorem.

An important example of a $3$-form is the \textit{magnetic helicity} $h_m$ \cite{Enk13},
defined as an integral over all space:
\begin{equation}
h_m = \int d^{3}r\, \A\bdot \B.\label{3.75}
\end{equation}
To make the directed volume element and the vector potential explicit, we write 
 \begin{equation}
\A\bdot(\boldsymbol{\nabla}\boldsymbol{\times} \A )= 
\A\bdot(-i\boldsymbol{\nabla}\boldsymbol{\wedge} \A )=
-i(\A\boldsymbol{\wedge}\boldsymbol{\nabla}\boldsymbol{\wedge} \A) . 
 \label{3.76}
 \end{equation}
Whence,
\begin{equation}
h_m =\int d^{3}\r\bdot
(\A\boldsymbol{\wedge}\boldsymbol{\nabla}\boldsymbol{\wedge} \A )^{\dagger} .
 \label{3.77}
 \end{equation}
Ra\~nada \cite{Ranada92,Ranada14} recognized that the $h_m$ can be identified with the 
 \textit{Hopf index} for a multi-valued vector potential with integer values.
In Section IV we propose further to identify the Hopf index with the principal quantum number in atomic physics.

 \section{ Dirac Theory of the electron clock}\label{sec:III}

This Section reviews key features of\emph{ Dirac electron theory} developed in the
preceding paper \cite{Hest19a}
to prepare for extension to a unified\emph{ Maxwell-Dirac Theory} of the electromagnetic vacuum in the following Section.  
Then we discusses empirical evidence the that the electron is a quantum oscillator that serves as a digital clock in physical processes.
Finally, the electron's  
anomalous magnetic moment is explained and calculated to seven significant figures.

\subsection{Local observables on Dirac streamlines}

We have seen that, with \emph{Spacetime Algebra}, the standard matrix version of the Dirac equation can be reformulated as a 
\emph{real Dirac equation}:  
\begin{equation}
\square\Psi \i\hbar-\frac{e }{c}A\Psi=m_{e}c\Psi\gamma_0\,.\label{4.1}
\end{equation}
Here $A=A(x)=A_\mu\gamma^\mu$ is the electromagnetic vector potential for external sources and the unit bivector $\i=i\bsig_3$ encodes the crucial property of spin.
The spinor ``\emph{wave function}" $\Psi=\Psi(x)$ has the \emph{Lorentz invariant} decomposition
\begin{equation}
\Psi=\rho^{\smallhalf} Re^{i\beta},\label{4.2}
\end{equation}
where $\rho =\rho(x)$ is scalar-valued density and ``\emph{rotor}" $R= R(x) $ is normalized to $R\tR= \tR R=1.$
The rotor $R$ has a unique decomposition into the product
\begin{equation}
R=VUe^{-i\bsig_3\varphi } ,  
\label{4.3}
\end{equation} 
where rotor $V=(v\gamma_0)^\half $ defines a boost to the electron's center of mass, the spatial rotor
\begin{equation}
U=U_1U_2=e^{-i\bsig_3\varphi_1 }e^{-i\bsig_1\varphi_2 }  
\label{4.4}
\end{equation}
describes kinematics of spin, and $\varphi = \varphi(x)$ is identified as the quantum mechanical phase of the wave function.

The Dirac wave function  determines a timelike velocity $v=v(x)$ defined by
\begin{equation}
\Psi\gamma_0\widetilde{\Psi}=\rho R\gamma_0\tR=\rho V\gamma_0\tV=\rho v,\label{4.5}
\end{equation}
and the Dirac equation implies  this is a conserved vector current with
\begin{equation}
\square \cdot (\rho v) =0.\label{4.6}
\end{equation}
The \textit{Dirac current} fills a given region of spacetime with
a congruence of non-intersecting timelike \textit{paths}.
Every path (\textit{streamline} or \textit{fiber}) $x(\tau)$ has a velocity $v=\dot{x}= e_0(x(\tau))$ and a comoving frame of ``local observables''
\begin{equation}
e_\mu=e_\mu(x)=e_\mu(x(\tau))\label{4.7}
\end{equation}
with $\mu=0,1,2,3$.
The vector 
\begin{equation}
s=\half \hbar\, e_0=\half \hbar\, e_0(x(\tau))\label{4.8}
\end{equation}
is identified with the electron spin vector on the path,
while the spin bivector
\begin{equation}
S=isv=\half \hbar\, e_2 e_1\label{4.9}
\end{equation}
specifies the plane of electron phase $\varphi=\varphi(x(\tau))$. (See Fig.1).
This characterization of local observables in Dirac theory is completely general, without any approximation, so it is equally applicable to Copenhagen or Pilot Wave 
interpretations of QM.
Differences arise in the physical interpretation ascribed to electron paths.

\begin{figure}
\centering
\includegraphics[width=0.7\linewidth]{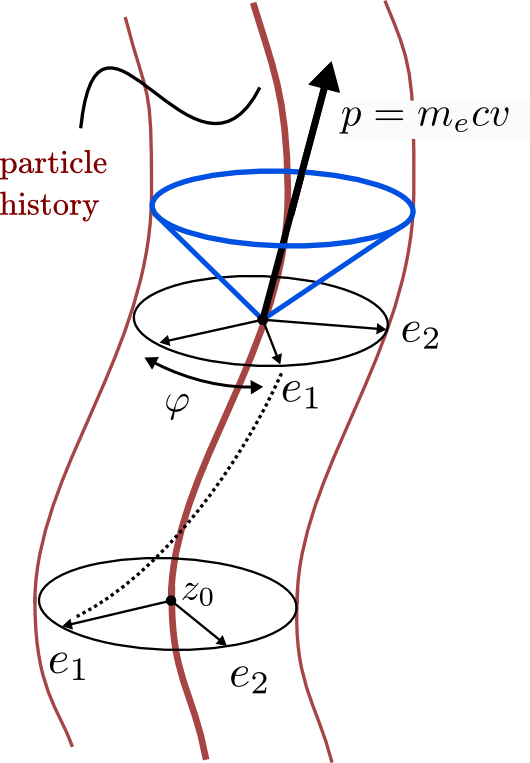}
\caption{The ``spinning frame'' of local observables along an electron path is depicted in a direction orthogonal to the spin vector.}
\label{fig:zitterSolution1r1}
\end{figure}

\subsection{Pilot particle and electron clock}

Physical implications of the Dirac wave function $\Psi(x)$ are studied in 
\cite{Hest19a} by applying conservation laws of momentum and angular momentum to construct equations of motion for local observables on a particle path. One outcome is the \textit{Pilot Particle Model} (PPM), which provides a new approach to studying the Hydrogen atom.

Another outcome is the \textit{Zitter Particle Model} (ZPM), which 
requires a subtle modification of Dirac equation so the electron path is a lightlike helix.
The ZPM provides the core for electron theory in all that follows.
As explained in \cite{Hest19a},
the ZPM regards the worldline of the electron as a lightlike circular helix $z_e=z_e(\tau)$  
with velocity $u=\dot{z_e}$ (Fig.2).
The helix is centered on a timelike path  $z=z(\tau)$
with velocity $v=\dot{z}$ normalized  to $v^2 =1$. This sets a time scale for the proper time parameter $\tau.$ 
A length scale is set by identifying the radius of the helical path with   
\begin{equation}
|z_e(\tau)-z(\tau)|=\lambda_e=c/\omega_e,\label{7.1}
\end{equation}
where $2\lambda_e=\hbar/m_e c$ is the reduced Compton wavelength and $\omega_e$ is the frequency of the circular motion called \textit{zitter}.
We refer to the point $z_e(\tau)$ as the \textit{center of charge} (CC) and to $z(\tau)$ as the \textit{center of mass} (CM).

As first proposed in \cite{Hest74a,Hest74b} and extended to lightlike particles in \cite{Hest90,Hest10}, we define a comoving frame of \textit{local observables} attached to the CM by 
  \begin{equation}
 e_{\mu}(\tau)=R\gamma_{\mu}\tR, \label{7.2}
 \end{equation}
where rotor $R = R(\tau)$ is normalized to $R\tR =1.$
Then we identify $e_0 =v$ and define $r_e =z_e-z=\lambda_e e_1$ as the radius vector for the zitter.  Thus we can regard $e_1$ as the ``hand of the electron clock."
  
We identify the unit vector $e_{2}$ as the tangential velocity of the zitter. There are two distinct senses for the circulation, which we identify with the   \textit{electron/positron} distinction called 
``\emph{chirality}.''
Accordingly, 
we have two null vector  tangential velocities:
\begin{equation}
e_{\pm}= v\pm e_{2}=R\gamma_{\pm}\tR,\quad\hbox{with}\quad\gamma_{\pm}
=\gamma_{0}\pm\gamma_{2}.\label{7.3}
\end{equation}
Unless otherwise noted, we restrict our attention to the electron case here, and our choice of sign for the chirality is in agreement with \cite{Hest10}.  
Accordingly, we define the electron ``\emph{chiral velocity}" $u$  by
\begin{equation}
u=R\gamma_{+}\tR=v+e_{2}. \label{7.4}
\end{equation}
The \textit{rotational velocity} of the zitter is a spacelike bivector defined by
\begin{equation}
\Omega =2\dot{R}\tR, .\label{7.5}
\end{equation}
so that
\begin{equation}
\dot{e}_\mu=\Omega \bdot e_\mu .\label{7.6}
\end{equation}
In particular, 
\begin{equation}
\dot{r}_e=\Omega \bdot r_e=\lambda_e \Omega\bdot e_1  =(e_1 e_2)\bdot e_1=e_2  ,\label{7.7}
\end{equation}
in agreement with (\ref{7.4}).

\begin{figure}
\centering
\includegraphics[width=0.7\linewidth]{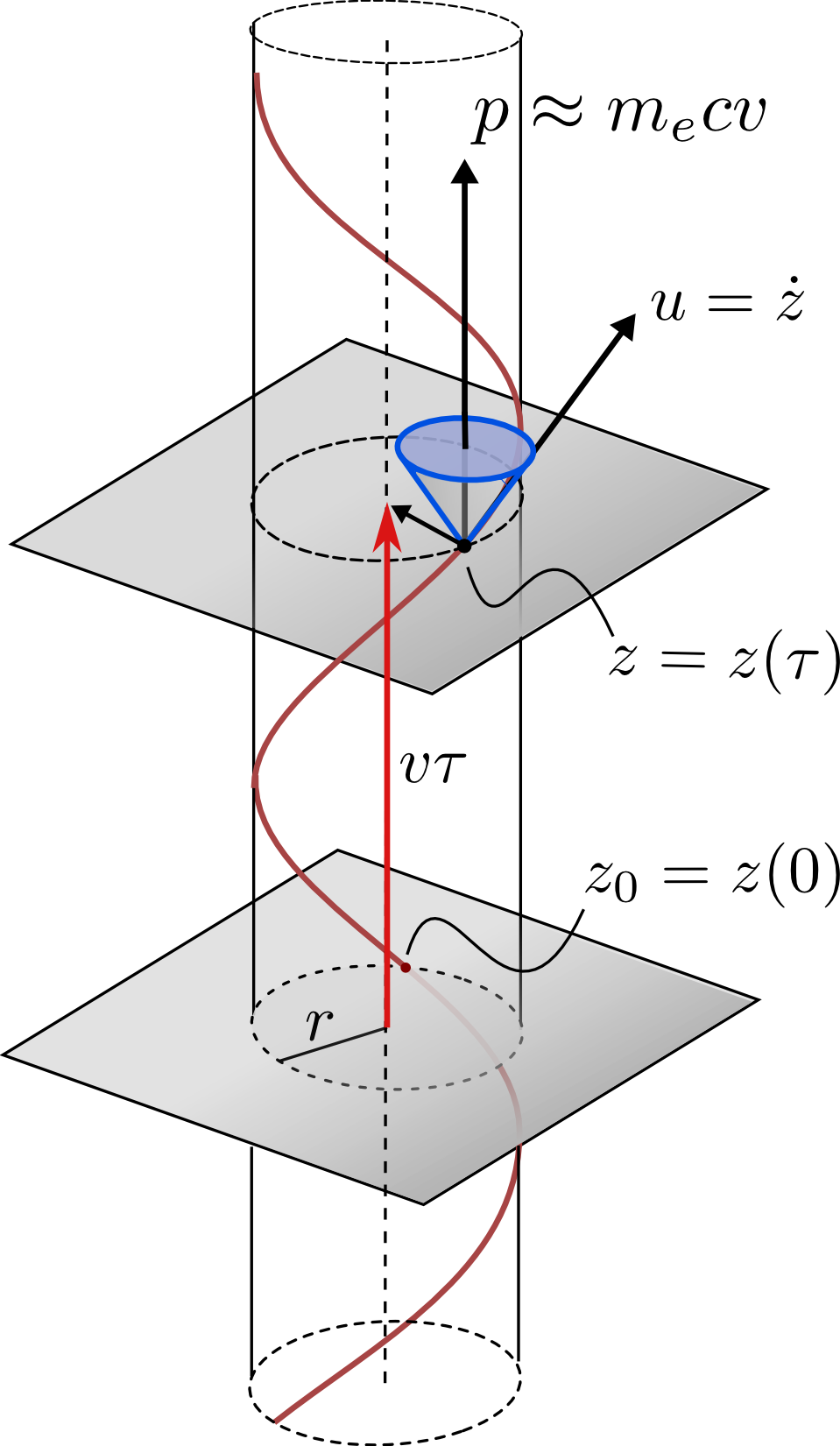}
\caption{ In the Zitter Particle Model the electron path is an oriented lightlike helix with an opposite orientation for a positron.
}
\label{fig:zitterSolutionr1}
\end{figure}

As shown in \cite{Hest19a},
zitter mechanics  can be described in terms of mass current $m_ev$, momentum $p$ and spin angular momentum $S$.  The 
\emph{chiral} spin bivector $S$  can be expressed   in  several equivalent forms:
\begin{equation}
S=ud=v(d+is)=ius,\label{7.8}
\end{equation}
Note that the null velocity $u^{2}=0$ implies null spin bivector $S^{2}=0$.
It follows that the free particle zitter motion can be reduced to a single equation:  
\begin{equation}
\Omega S= pv.\label{7.9}
\end{equation}
The bivector part of this expression gives us the spin equation of motion:
\begin{equation}
\dot{S}=\Omega\boldsymbol{\times} S= p\wedge v.\label{7.10}
\end{equation}
And the scalar part  gives us
\begin{equation}
m_ec=p\bdot v=\pm\Omega\bdot S ,\label{7.11}
\end{equation}
where the sign distinguishes electron/positron chirality.

Note that these equations are consistent with identifying momentum $p$ with either the timelike vector $m_ecv$ or the null vector $m_ecu$. This ambiguity is resolved in the next Section, where $eu$ is identified as a charge current. On the other hand, the opposite choice was tacitly introduced in 
 \cite{Hest19a} by dividing
(\ref{7.10}) with $v$ to get 
\begin{equation}
p=m_ecv+\dot{S}\bdot v .\label{7.12}
\end{equation}
Of course, we could not instead divide  by the null vector $u$.  
The fact that introduction of electron charge seems to be needed to resolve this ambiguity about mass may be an important clue about the relation to charge and mass in the theory.

 \subsection{Electron clock and fundamental constants}

The \textit{Josephson effect} gives very accurate data for the quantum flux unit $h/e$, while the quantum Hall effect gives very accurate data for the \textit{Hall impedance}
$h/e^2$. Together they measure the fundamental constants $[h]$ and $[e]$  to an accuracy approaching 9 decimal places.
As argued by Post \cite{Post99},
this provides ample reason to replace the standard system of units $[l, t, m, q]$ for [length, time, mass, charge] by a system $[l, t, h, e]$ with \textit{metric} units $[l,t]$ for length and time and \textit{topological} units $[h,e]$ for action and charge with integer values.
Moreover, it supports identification of the electron clock as the fundamental mechanism grounding the units of action and charge.

The \textit{zitter model} of the electron clock
has been memorably described by Consa 
\cite{Consa18}
 as a superconducting LC circuit composed of two indivisible elements: ``\textit{a quantum of electric charge and a quantum of magnetic flux, the product of which is equal to Planck's constant. The electron's magnetic flux is simultaneously the cause and the consequence of the circular motion of the electric charge:}''
\begin{equation}
  e\Phi_e=h. \label{7.25} 
\end{equation}
Motion of the charge causes a frequency dependent electric current $I=ef_e$ and an electric voltage $V_e=hf_e/e$ with a fixed ratio given by the Hall impedance:  
\begin{equation}
  Z_e=\frac{V_e}{I_e}
  =\frac{hf_e/e}{ef_e}
  =\frac{h}{e^2}. \label{7.251} 
\end{equation}
The electron's circular orbit at the speed of light constitutes the most elementary superconductor possible, generating the flux quantum  
\begin{equation}
\Phi_e=V_eT_e
=\frac{hf_e}{e}\frac{1}{f_e}
=\frac{h}{e}. \label{7.252} 
 \end{equation}
 Analysis of the quantum LC circuit is straightforward with standard techniques. The Capacitance (C) and Self Inductance (L) are defined by
\begin{equation}
   L_e=\frac{\Phi_e}{I_e}
   =\frac{h}{e^2 f_e},
   \label{7.253} 
 \end{equation}
\begin{equation}
   C_e=\frac{e}{V_e}
   =\frac{e^2}{h f_e}.
   \label{7.254} 
 \end{equation} 
Applying the formulas for an LC circuit, we get expressions for the impedance and natural frequency in the form
\begin{equation}
   Z_e=\sqrt{\frac{L_e}{C_e}}
   =\frac{h}{e^2},
   \label{7.255} 
 \end{equation}
\begin{equation}
   f_e =\frac{1}{\sqrt{L_eC_e}} =f_e.
   \label{7.256}   
 \end{equation}
While the energy of the particle oscillates between electric and magnetic energy, the average energy has the constant value
\begin{equation}
   E=\frac{LI^2 }{2}+
   \frac{CV^2 }{2}
   =\frac{hf }{2}+
      \frac{hf }{2}=hf=\hbar\omega.
   \label{7.257} 
 \end{equation} 
As Consa observes, the above calculations are valid for any elementary particle with charge $e$, vibration frequency $f$, and energy given by Planck's equation $E=hf=\hbar\omega$.

\subsection{Electron anomalous magnetic moment}
  
The basic free particle solution of the zitter equations of motion is specified by a constant spacelike bivector 
\begin{equation}
 \Omega =\omega_e e_2e_1, 
 \label{7.13}
 \end{equation}
 where $\omega_e=2m_e c^2/\hbar$ is the zitter frequency.
 There is, however, a more subtle solution that was recognized only recently by Oliver Consa 
 \cite{Consa18}, who also fully recognized its extraordinary significance in explaining the intrinsic anomalous magnetic moment of the electron.

When dealing with free particle solutions, it is convenient to work 
with the spacetime split of the comoving frame  (\ref{7.2}) defined by $v=e_0$ to introduce a  frame of relative vectors 
 \begin{equation}
 \e_k=\e_k(\tau)=U\bsig_{k}
 U^{\dagger},
 \label{7.14}
 \end{equation}
 so that
 \begin{equation}
 \dot{\e}_k=\boldsymbol{\omega}\boldsymbol{\times}\e_k, \quad \hbox{where} \quad -i \boldsymbol{\omega}=2\dot{U}U^{\dagger}
 \label{7.15}
 \end{equation}
 is the angular velocity.
 In the simplest solution, called \textit{circular zitter}, the particle orbit is generated by  rotating vector $\r_e=\lambda_e \e_1$ with period $\tau_e= 2\pi /\omega_e$.
 
In the new solution, called \textit{toroidal zitter}, the electron's circular orbit precesses around a fixed axis in a plane inclined at a fixed angle specified by the anomalous magnetic moment. As the orbit precesses, it sweeps out a toroidal surface as depicted in Fig.3. The precessing motion generates two orthogonal electric currents, as indicated in Fig.3. The first is a circular current with radius $r_1$. The second is a \textit{poloidal current} that circulates around the poloidal cross section with radius $r_2$. 

In the zitter model the electron magnetic moment  is generated by the 
charge current circulating in electron zitter.
The poloidal component of this current cancels out, thus reducing the net magnetic moment. Such cancellation of poloidal currents in a solenoid is well known in the design of electrical antennas. Here it explains the origin of the electron's anomalous magnetic moment. Next we show how to calculate its value.

\begin{figure}
\centering \includegraphics[width=0.7\linewidth]{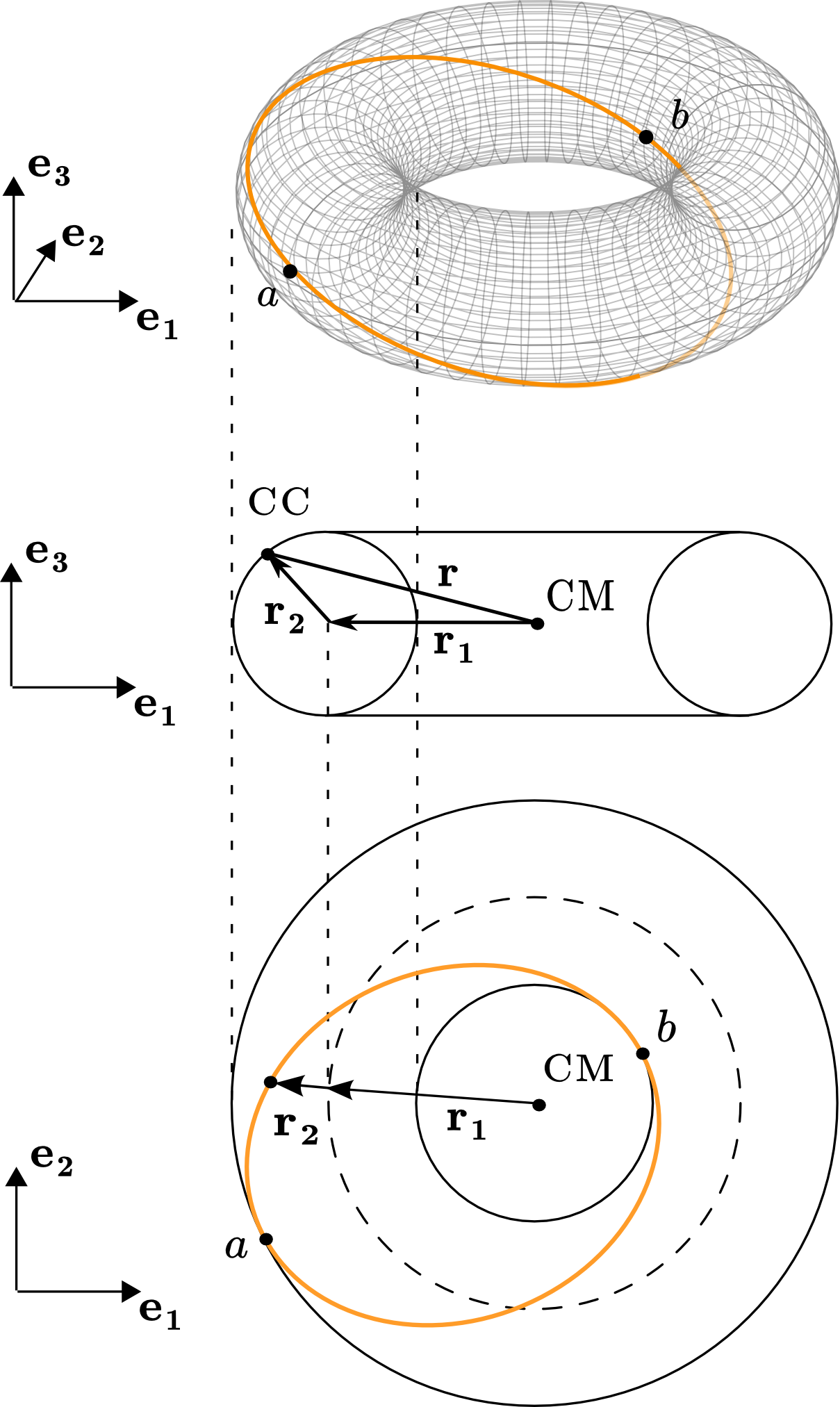}
\caption{ In a plane rotating with angular velocity $\omega_1 \bsig_{3}$ around the CM,
 the electron's lightlike circular orbit lies on a toroidal surface called the \textit{electron energy shell} that projects to the orbital plane as an ellipse with antipodes $a$ and $b$.}
\label{fig:torus cross section} \end{figure}

We consider a particle orbit $\r=\r(\tau)$  composed of a pair of orthogonal rotating vectors
\begin{equation}
 \r =\r_1 +\r_2 = r_1\e_1 +r_2\e_3,  
 \label{7.16}
\end{equation}
so the orbit circulates at the speed of light on a torus with radius $r_2$.
Depiction of this path in a frame rotating with angular velocity $\omega_1 \bsig_{3}$ around the CM is given in Fig.3.

The toroidal path can by parametrized with angles $\theta$ and $\phi$ as follows. 
First,  describe a circle in the $\bsig_1\bsig_3=-i\bsig_2$ plane displaced along a line in direction $\bsig_1$ by
\begin{align}
\r_0(\phi)
&=(r_1 +r_2 e^{-i\bsig_2\phi})\bsig_1 \notag\\
&=r_1\bsig_1 +r_2U_2\bsig_1  U_2^\dagger. 
\label{7.17}
\end{align} 
Then  rotate the circle around the $\bsig_3$ axis to construct a torus given by
\begin{equation}
 \r(\theta,\phi) =U_1\r_0(\phi) U_1^\dagger
 \label{7.18}
\end{equation} 
and generated by rotors
\begin{equation}
U_1=e^{-i\bsig_3\theta/2}
\quad \hbox{and}\quad
U_2=e^{-i\bsig_2\phi/2}
\label{7.19}.
\end{equation}
A closed curve $\r(\tau)$ with constant frequencies 
 is generated by setting 
$ \theta=\omega_1\tau$ and $ \phi=\omega_2\tau$.
The rotors combine to define a single spinor generating toroidal zitter:
\begin{equation}
U=U_1U_2=e^{-i\boldsymbol{\omega}_1\tau/2}e^{-i\boldsymbol{\omega}_2\tau/2}=U(\tau), \label{7.26}
\end{equation}
with $\boldsymbol{\omega}_1 =\omega_1\bsig_3$ and  $\boldsymbol{\omega}_2 =\omega_2\bsig_2$.
This differentiates to 
\begin{equation}
\dot{U}=\dot{U}_1U_2+U_1\dot{U}_2=-\half[i\boldsymbol{\omega}_1U+Ui\boldsymbol{\omega}_2],   \label{7.27}
\end{equation}
which, in accord with  (\ref{7.15}), gives us an explicit expression for the rotational velocity of the motion:
\begin{align}
 2\dot{U}U^{-1}&=-i[\boldsymbol{\omega}_1+U_1\boldsymbol{\omega}_2U_1^{\dagger}] \notag\\&=-i[\boldsymbol{\omega}_1+\boldsymbol{\omega}_2\,e^{i\boldsymbol{\omega}_1\tau}]=-i\boldsymbol{\omega}. \label{7.29}
\end{align} 
Whence, the particle velocity is given by $\dot{\r}=\boldsymbol{\omega}\boldsymbol{\times}\r$.

Since the electron circulates at the speed of light, we have
\begin{equation}
 \dot{\r}^2 =\Big(\frac{d\r}{d\tau}\Big)^2
 = -(\r\wedge\boldsymbol{\omega})^2
 =\r^2\boldsymbol{\omega}^2
 -(\r\bdot\boldsymbol{\omega})^2
 = c^2,  
 \label{7.20}
 \end{equation}
where $\r\bdot\boldsymbol{\omega}=0$ for a free particle with the CM at rest.  
For an orbit precessing about the CM as depicted in Fig.3,
the average velocity is reduced to $c/g$  by a factor $g$ called the
 \textit{helical $g$-factor} by Consa. Accordingly, the radius $r=|\r|$ projects to a mean radius $r_1$ with mean velocity $\omega r_1=c/g$, and
(\ref{7.20}) gives us
\begin{equation}
\frac{\r^2\boldsymbol{\omega}^2}{c^2}=\bigg(\frac{r}{gr_1}\bigg)^2. 
 \label{7.21}
 \end{equation}
Solving for $g$ we get
\begin{equation}
g^2=1+(r_2/r_1)^2 
+2\r_2\bdot \r_1^ {-1},
 \label{7.22}
 \end{equation}
and averaged over one zitter cycle, the cross term gives
 \begin{equation}
\overline{2\,\r_2\bdot \r_1^{-1}}
=(r_2/r_1)^2\,\overline{\cos\theta}=\half (r_2/r_1)^2
 \label{7.23}
\end{equation} 
with the combined result   
\begin{equation} g=\sqrt{1+(3/2)(r_2/r_1)^2}. \label{7.24} \end{equation}
Then Consa evaluated this quantity by comparing it with Schwinger's celebrated result from quantum electrodynamics:
\begin{equation}
  g(QED)=1+\alpha/2\pi =1.0011614,
  \label{7.24a}
\end{equation}
where $g/2$ corresponds to the standard definition for the
 \textit{anomalous $g$-factor} $a_0$.

Accordingly, by expanding the radical in  (\ref{7.24}) to first order we get
\begin{equation}
g-1=(3/4)(r_2/r_1)^2 
=\frac{\alpha}{2\pi}.
 \label{7.24b}
 \end{equation}
The right side of this expression gives $g$ a quantitative value
while Fig.3 gives it a 
\textit{physical interpretation}, where $r_1$ and $r_2$ are outer and inner radii of an electron energy shell.
The outer radius is a known quantity, the  \textit{zitter radius}   $\lambda_e=r_1$, while the inner radius $r_2 $ is a free parameter in the model.

According to (\ref{7.11}), the energy in free electron zitter is given by $m_ec=\Omega\bdot S=\boldsymbol{\omega}\bdot \s$, so with constant spin $S=i\s$, the $g$-factor describes a zitter frequency shift due to free precession producing an energy (mass) shift given by
\begin{equation}
(m_e+\Delta m_e)c=\Omega\bdot S. \label{7.24bb}
\end{equation}
The same result has been found by others \cite{Muralidhar14}.

Conventional QED provides no physical explanation whatsoever for the Schwinger result (\ref{7.24a}), despite its mathematical sophistication and claims for unprecedented accuracy. Indeed, I heard Feynman himself declare that such lack of physical insight would not be tolerated in any other branch of physics.

But now, with (\ref{7.24b}) and (\ref{7.24bb}), we have fresh theoretical insight explaining \textit{the electron anomalous magnetic moment as free precession of electron zitter}, quite analogous to free precession of a rigid body in classical mechanics. That should drive a new round of research into its implications.

In the meantime, there is plenty of room for design of experiments to probe related issues, especially on the role of magnetic resonance. For example:

Rotor methods for magnetic resonance measurement are given in \cite{Hest86}, along with many other coordinate-free
applications to rotational dynamics. 
Though a properly tuned magnetic resonance measurement may activate toroidal zitter in the electron, external fields are not necessary to maintain it. 

If the toroidal zitter can be quenched or activated at will, then the electron has at least two distinguishable internal states and might thereby  serve as the ultimate magnetic storage device. 
On the other hand, if toroidal zitter is a persistent intrinsic property of the electron, then circular zitter should simply be regarded as an approximation.

Such issues about toroidal zitter are likely to be central to elementary particle theory as proposed in Section VI.


\section{ Maxwell--Dirac Theory}\label{sec:IV}

 \textit{Born-Dirac} theory supports the
``Pilot--Wave'' interpretation of quantum mechanics originally proposed by de Broglie \cite{Hest19a}. 
But that is only half of de Broglie's proposal,
which he called  \textit{double solution} theory \cite{Broglie87}.
In consonance with his realist perspective on quantum mechanics, he proposed that there must be two distinct kinds of solution to the wave equation. Besides the pilot wave, there must be some kind of {\it singular solution} describing a real particle without involving probability.
In other words, he proposed a unique kind of \textit{wave--particle duality} that might better be called \textit{field--particle duality}.

De Broglie insisted that relativity is an essential ingredient of fundamental quantum mechanics, and he noted that monochromatic plane wave solutions of the relativistic wave equation determine well--defined particle paths, just as we have seen for the Dirac equation in Born-Dirac theory. However, he never found a convincing way to define a singularity that represents a physical particle. In this section we introduce a new physical interpretation of the Dirac wave function 
$ \Psi $ that seems to do the trick.
\textit{The essential idea is to give back to the electron its charge and electromagnetic field}, which were ignored in the original neutered version of Dirac theory. This necessarily localizes the electron to the point source of its Coulomb field. The trick is to do it in a way that is consistent with well established features of Dirac theory. 

According to the Born rule, the Dirac current $\Psi\gamma_{0}\widetilde{\Psi}=\psi\gamma_{0}\tpsi=\rho v$ is to be interpreted as a probability current so the dimensionless function  $\rho=\rho(x)=\psi\tpsi$ must be a probability density; therefore
$e \rho(x)$ should be interpreted as a ``probable'' charge density.
In the early days of QED, Furry and Oppenheimer \cite{FurOpp34} called this proportionality into question by asserting that $e \rho(x)$ must be interpreted as a physically real charge density to enable comparison with ``real charges'' in classical electrodynamics that produce real electromagnetic fields.
Indeed, second quantization was soon invented to do that, but it involves some dislocation from the original Dirac theory. 
However, there is a subtle alternative to this approach that has not been heretofore considered.

Our formulation of Born-Dirac as a classical field theory in \cite{Hest19a}  facilitates comparison with Maxwell's electromagnetics.
To coordinate the two to produce a fully integrated Maxwell-Dirac theory, we need some educated guesses to guide us.
To that end, I propose three fundamental anzatz's,

First, we invoke \textit{``de Broglie's clock ansatz''} already articulated in the ZPM model of the electron clock  in the preceding Section.
 
Second, we note from Section II that Blinder's assumption that \textit{charge density is proportional to mass density} in Maxwell theory can be carried over to Dirac theory where both quantities are associated with the Dirac current. Hence Blinder's argument relating electron mass to impedance of the vacuum should apply to the interpretation of the Dirac wave function in some way.
Let me call this the ``\textit{Blinder ansatz.}"

Third, we recall London's assumption that for electrons in a superconductor the magnetic vector potential is proportional to the charge current \cite{London50}. As all interactions in the Dirac equation are mediated by vector potentials, we look for a comparable relation to the electron's magnetic vector potential.
Let's call that the ``\textit{London ansatz.}"

Now, to incorporate zitter into the Dirac equation 
in accord with \textit{``de Broglie's clock ansatz,''} we simply replace the timelike velocity $v=R\gamma_0\tR$ in the Dirac current (\ref{4.6}) with the lightlike velocity of the electron's helical path:      
\begin{equation}
u=R(\gamma_0+\gamma_2)\tR=e_0+e_2=\dot{z}_e(\tau),\label{4.10}
\end{equation}
This decomposes the zitter into the timelike CM velocity
\begin{equation}
v=R\gamma_0\tR=\dot{z}(\tau) =\bar{u}(\tau)\label{4.11}
\end{equation}
and the fluctuating  spacelike velocity $e_2=R\gamma_2\tR$ of the circular zitter with $\bar{e}_2 =0$.
In a similar way we identify the spacelike spin bivector 
in (\ref{4.8}) as the zitter average $\bar{S}$ of the lightlike spin bivector (\ref{7.8}).

With the electron's helical path embedded in the wave function $\Psi(x)$, its functional form reduces to $\Psi(x-z_e(\tau))$
and factors into the product:
\begin{equation}
\Psi = VU\psi,\label{4.12}
\end{equation}
where zitter kinematics is incorporated in the rotor 
\begin{equation}
U=U(\tau)=U_1(\tau)U_2(\tau)\label{4.12a}
\end{equation}
with form specified by (\ref{7.19})
\begin{equation}
\psi = \rho^{1/2}e^{-\i\varphi }=e^{-(\alpha+i\bsig_3\varphi) }.\label{4.13}
\end{equation}
has the familiar form for a Schr\"odinger wave function, but, of course, with an imaginary unit specified by the bivector $\i=i\bsig_3$. 

A truly remarkable implication of the wave function $\psi=\psi(x-\dot{z}(\tau))$ given by (\ref{4.13}) is its reduction of coupling to the electromagnetic vacuum
to a pair of retarded electric and magnetic scalar functions $\alpha(x-\dot{z}(\tau))$ and $\varphi(x-\dot{z}(\tau)).$ Details are provided by the complementary ansatzes of Blinder and London.

The \textit{Blinder ansatz} identifies the Dirac current $ m_{e}c\rho v $  with the Blinder form for the retarded potential of a point charge (\ref{3.15}) by assuming that the Dirac \textit{density} $ \rho=\rho(x) $ is reciprocal to tne impedance $\epsilon $ of the  vacuum. Thus, it holds that 
\begin{equation}
\rho=\epsilon^{-1}=e^{-\alpha} \quad \hbox{where}\quad \alpha = \lambda_{c}/r,\label{4.14}
\end{equation}
while the \textit{classical electron radius}
$ \lambda_{c}= e^{2}/m_{e}c^{2} $ serves as a charge/mass scaling length and $ r=(x-z_e(\tau))\bdot v $ is the\textit{ classical retarded distance} from a point singularity at the position $z_e(\tau)$ of the electron. In other words, we identify
\begin{equation}
\frac{e}{c}A_{c}\equiv\frac{e^{2}}
{c\lambda_{c}\epsilon}u=m_{e}c\rho u,\label{4.15}
\end{equation}
or, more simply,
\begin{equation}
\lambda_{c}A_{c}= e\rho u,
\label{4.16}
\end{equation}
with the classical ``\textit{Coulomb vector potential}"  $ A_{c} $ for a point charge.
Acceptance of this ``Blinder ansatz'' commits us to solutions of the Dirac equation as a function of retarded position.
Note that the source CC of the electric field $z_e(\tau)$ is displaced from the CM $z(\tau)$ by the radius vector 
\begin{equation}
r_e=z_e(\tau)-z(\tau)=\lambda_e e_1,
\label{4.17}
\end{equation}
where $e_1$ is the electron ``clock vector''
and $\lambda_e=\hbar/2m_e c$ is the zitter radius.

A crucial feature of the Blinder function is that  $\rho(z_e(\tau))=0$ everywhere along the electron path $z_e(\tau)$, and thus at a single point on any 3-D spacelike hypersurface.
At that point the phase in the wave function  is undetermined, so it can be multivalued.
This mechanism can also serve to pick out a particle path in  Pilot Wave theory.

If indeed the electron's Coulomb potential is already inherent in the Dirac equation as the Blinder ansatz requires, then we should expect to find the electron's magnetic potential there as well.
This leads us to examine the Gordon current, where we note that it includes a ``magnetization current.'' Proposing this as a specific realization of the ``\emph{London ansatz},'' we reinterpret that term as a ``magnetic potential.''  Thus, in analogy with the Blinder potential (\ref{4.16}) we write 
\begin{equation}
\frac{e}{c}A_{m}\cong-\square\bdot
(\rho S).\label{4.18}
\end{equation}
The congruence sign $\cong$ serves to indicate that the two quantities are mathematically equivalent but have different physical interpretations. In Born-Dirac theory presented in \cite{Hest19a} the spin density $\rho S$ represents a distribution of spins associated with distinct particle paths. As will become evident, in the present Maxwell-Dirac theory the analogous quantity represents a density of magnetic field lines.
Combining the electron's electric and magnetic vector potentials, we have a complete analogy with the entire Gordon current:
\begin{equation}
\frac{e}{c}A_{e}=\frac{e}{c}(A_{c}+A_{m})\cong m_{e}c\rho u-\square\bdot(\rho S).\label{4.19}
\end{equation}
According to (\ref{4.16}), therefore, this reduces the Dirac equation to:
\begin{equation}
\rho p=\rho(P-\frac{e}{c}A)=\frac{e}{c}A_{e}.
\label{4.20}
\end{equation}
This equation describes a remarkable duality between the charge current along the electron path and the electromagnetic field it generates and propagates by Maxwell's equation:
\begin{equation}
\square(\rho^{-1}\square A_{e})=0.\label{F.3}
\end{equation}
Recall that the Blinder function for $\rho$, given by (\ref{4.14}), vanishes on the electron path, so the zero can be factored out of the momentum balance equation (\ref{4.20}).

With (\ref{4.20}) we have boiled down the Dirac equation to a relation between the electron's vector potential $A_{e}$ and what we can now identify as an energymomentum density of the vacuum $\rho P$.
We note that, as an equation for momentum balance, it comes close to the ideal of putting the electron vector potential $ A_{e} $ on equal footing with the vector potential $ A $ for external interactions. The only difference is that the density $ \rho $ is an essential part of  $ A_{e} $ but not of  $ A $. Later on, we take this as a clue to a many particle generalization.
For present purposes, we can ignore external interactions, though we maintain the electron's interaction with the vacuum.

Electric and magnetic components of the electron potential $A_e$ are separated by a spacetime split with respect to the CM velocity $v$; thus 
\begin{equation}
  A_{e}v = A_{c}\bdot v + A_{m}\wedge v= e\rho/\lambda_{c} +\A_{m} .\label{4.21}
\end{equation}
Then a $v$-split of the canonical momentum $ P $ gives us 
\begin{equation}
Pv = P\bdot v + P\wedge v= P_{0}+\mathbf{P}
  =m_{e}c+\hbar\boldsymbol{\nabla}\varphi .\label{4.22}
\end{equation}
Hence, electron energy is identified with the electric potential
\begin{equation}
  P_{0} =m_{e}c=\frac{e}{c}A_{c}\bdot v \rho^{-1}, \label{4.23}
\end{equation}
while  momentum is identified with the magnetic potential:
\begin{equation}
\mathbf{P}=\hbar\boldsymbol{\nabla}\varphi=\frac{e}{c}\A_{m} \rho^{-1} .\label{4.24}
\end{equation}
Note that, though electric and magnetic fields  are separable in  the Gordon current (\ref{4.19}), they are not independent.
They are intimately coupled by the  vacuum density $\rho=\rho(x)$, as we now show.

Regarding the zitter average of spin $\s$ as constant, we  calculate the magnetic vector potential from:
\begin{align}
 \A_{m}\rho^{-1} &=\frac{-c}{e\rho}\boldsymbol{\nabla}
  \bdot (\rho i \s) 
  =\frac{c}{e\rho}\boldsymbol{\nabla}
    \boldsymbol{\times} (\rho  \s)\notag\\
&=\frac{c }{e} \boldsymbol{\nabla}\boldsymbol{\times}\left(\frac{-\lambda_{c}\s}{r}\right) = g_{e}\frac{\r\boldsymbol{\times}\s}{r^{3}},\label{4.25}
\end{align}
which is the familiar potential for a magnetic dipole,
where $\lambda_{c}=e^2/m_e c^2$ is the ``classical electron radius'' and
\begin{equation}
  g_{e} = c\lambda_{c}/e=e/m_{e}c  \label{4.26}
\end{equation}
is the correct $ g$-factor for a Dirac electron. What a surprise! For $g_{e}$ has been derived here from a  singular wave function  unknown in conventional Dirac theory. 
Indeed, conventional free particle solutions do not even have a magnetic  moment! Moreover, the derivation connects the magnetic vector potential  to the Coulomb potential, as expected in an integrated model of electric and magnetic forces in an electron.

Comparison of equation (\ref{4.24}) with
(\ref{4.25}) shows that $\rho$ serves as a Lagrange multiplier relating gradient to curl in the form
\begin{equation}
\hbar\boldsymbol{\nabla}\varphi=\boldsymbol{\nabla}\boldsymbol{\times}\boldsymbol{\pi} .\label{4.27}
\end{equation}
The general question of when the curl of a vector field $\boldsymbol{\pi}$ is equivalent to the gradient of a scalar has been studied by Kleinert
\cite{Kleinert08}. He shows that is possible when the scalar function is the projection of an area integral -- in physical terms, a flux generated by a current loop. Moreover, in general, the flux is multivalued, so we can conclude that 
\begin{equation}
\frac{e}{c} \oint d\x\bdot\A_{m}\rho^{-1}
=\oint d\x\bdot \mathbf{P}=
\hbar \oint d\varphi =nh,    \label{4.28}
\end{equation}
This assigns a definite quantum of flux $\Phi=\hbar/e$ 				to the electron's magnetic field, as anticipated by London and evidently measured in the quantum Hall effect \cite{Post99}.
In (\ref{3.77}) we identified it with the Hopf index and the principal quantum number in atomic physics.
It can be regarded as one more prediction of Dirac theory, provided the above identification of the electron vector potential with the Gorden current is accepted. In that case, \textit{the flux quantum ranks with electric charge as a fundamental property of the electron}, as, indeed, many since London have suggested. More precisely, it supports interpreting Dirac theory as modeling the electron as a fundamental singularity of the vacuum.


\subsection{Electron Self-Energy and Zitter}

Zitterbewegung is said to contribute to electron self-energy in QED, though that can be questioned because the integrals are divergent and must be removed by renormalization.
Weisskopf \cite{Weisskopf39} was the first to discuss the role of zitterbewegung in QED explicitly. Expressed in our lingo, he argues that zitter generates a fluctuating electric field. But when he calculates the zitter contribution to the energy in the field he gets a divergent result. In contrast, we show here that calculation with the zitter model is not only simpler, but the result is finite and equal to the expected result $m_{e}c^{2}$.
This is one reason to suspect that the zitter model may generate finite results in QED.

According to the \textit{Blinder ansatz}, 
the vacuum is characterized by 
the Blinder form  (\ref{4.14}) for the Dirac density $ \rho=\rho(x) $  given by 
\begin{equation}
\rho=\epsilon^{-1}=e^{-\lambda_{c}/r},\label{5.16}
\end{equation}
where the \textit{polarization radius}
$ \lambda_{c}= e^{2}/m_{e}c^{2} $ is a charge to mass scaling length, and 
\begin{equation}
r=(x-z_e(\tau))\bdot v\label{5.17}
\end{equation}
is the classical retarded distance from a point singularity at the position of the electron. 

To study zitter fluctuations, we shift electron position to the zitter CM.
That shift must be done in a way that preserves the "retarded distance" property of $r$.

With the backward Taylor expansion $z(\tau-\tau_{c})\approx z(\tau)+v(\tau)\tau_{c}$,
we get
\begin{equation}
z_e(\tau)=z(\tau-z_e)+r_{e}(\tau),\label{5.17a}
\end{equation}
hence
\begin{equation}
z_e=z(\tau)+v\tau_{c}+r_{e},\label{5.18}
\end{equation}
where all functions are evaluated at time $\tau$.
Requiring $[v\tau_{c}+r_{e}]^{2}= 0$, we find
\begin{equation}
\tau_{c}=\lambda_{e}/c\label{5.19}
\end{equation}
as the time for a light signal to propagate from the zitter center to the circulating particle. Accordingly, we have
\begin{equation}
z_e=z+v\tau_{c}+r_{e}=z+\lambda_{e}(v+e_{1}).\label{5.20}
\end{equation}
Then requiring $(x-z_e)^{2}=r^{2}-\r^{2}=0$,
for the the zitter \textit{retarded position} we get
\begin{equation}
\r=(x-z_e)\wedge v=\x-(\z+\r_e)\label{5.21}
\end{equation}
with\textit{ retarded distance}
\begin{equation}
r=|\r|=(x-z_e)\bdot v
=(x-z)\bdot v+\lambda_{c}.\label{5.22}
\end{equation}
This completes our characterization of the zitter vacuum.

Now, to incorporate the effect of zitter into the electron's electromagnetic field, we simply replace velocity $v$ with $u=v+e_{2}$ in (\ref{4.10}) to get a \textit{ Coulomb vector potential} of the form
\begin{equation}
\frac{e}{c}A_{c}\equiv\frac{e^{2}}
{ c\lambda_{c}}\frac{u}{\epsilon}=m_{e}c\rho u\label{5.23}
\end{equation}
To ascertain the implications of this change on the electron self energy, we need only consider how it modifies the Coulomb field of a free electron:
\begin{equation}
 F= \frac{m_{e}c^{2}}{e}(\square\rho)\wedge u=e\rho\,\frac{\hat{r}}{r^{2}}\wedge u= \E +i\B.\label{5.24}
 \end{equation}
 Note that $\hat{r}\bdot u= \hat{r}\bdot(v+e_{2})=0$, so we have $\hat{r}\wedge u = \hat{r} u$. Hence,
 \begin{equation}
   F^{2} = \mathbf{E}^{2}-\mathbf{B}^{2}+i
   \E\bdot\B=0.\label{5.25}
  \end{equation}
Since $ G=\rho^{-1}F = \mathbf{D}+i\mathbf{H}$, this implies that 
\begin{equation}
 \half\E\bdot\D=\half\B\bdot\mathbf{H}.\label{5.26}
  \end{equation}
Hence, the total energy in the field is
\begin{align}
W &=\half\int  (\E\bdot\D+\B\bdot \mathbf{H})
\, d^{3}r  \notag \\
&= \int  \E\bdot\D\, d^{3}r  =m_{e}c^{2},\label{5.27}
\end{align}
exactly twice the result obtained if the electron velocity were $v$ instead of $u$. As first suggested by Slater \cite{Slater26}, the reason for the difference is expressed by (\ref{5.26}), which tells us the  potential energy density of the circulating charge is equal to its kinetic energy density. However, this cannot be regarded as a fully satisfactory resolution of the notorious electron self-energy problem until the relation of gravitational to inertial mass is understood.  
  
More generally, we note that $\hat{r}uvu\hat{r}=uvu=2u$, hence 
  the field $F$ has an energymomentum density  
  \begin{equation}
   T(v) =\half Fv\,\tG= \frac{e^{2}\rho }{r^{4}}\,u,
     \label{5.28}
  \end{equation}
where $u=v+\dot{r}_{e}$ exhibits momentum fluctuations due to the rapidly rotating vector $\dot{r}_{e}$.
Note that these zitter fluctuations are not radiating, because the zitter velocity is orthogonal to the zitter radius vector.  
It remains to be seen if they can be identified with vacuum fluctuations of QED or the ubiquitous ground state oscillators proposed by Planck.

\subsection{Zitter fluctuations in QED}

Without trying to be comprehensive, let us consider how zitter fluctuations can account for important QED results.

The simplest case has already been considered
in Section III\,D, where the electron anomalous magnetic moment was explained as force free precession of the electron's zitter orbit, and described by a single rotor with form given by
(\ref{7.26}) as
\begin{equation}
U_g=e^{-i\boldsymbol{\omega}_1\tau/2}e^{-i\boldsymbol{\omega}_2\tau/2}. \label{5.29}
\end{equation}
Now note that his rotor can be incorporated into the electron wave function $\Psi$ by direct multiplication of the rotor part $U$ in (\ref{4.12a}) to get 
\begin{equation}
U'=U_g U.\label{5.30}
\end{equation}
The net effect on the electron equation of motion is to reduce the $g$-factor  $g_{e}$ specified in (\ref{4.26}) to
\begin{equation}
  g'_{e} = g_{e}/g,  \label{5.31}
\end{equation}
where the scaling factor $g$ is given by (\ref{7.24b}) and related equations.

To make another point,
it appears that zitter fluctuations will not alter the quantum conditions proposed for stationary states of a Pilot particle  as long as they are resonant with the quantum periods. Indeed, resonance of zitter fluctuations with orbital motion may be the most fundamental criterion for stationary states.

Though zitter will not alter quantum conditions, it should alter energy levels by smearing out the Coulomb potential over the zitter radius $\lambda_{e}$. This kind of explanation for the \textit{Lamb shift} was first proposed by Welton \cite{Welton48}.
Moreover, in $s$-states discussed in \cite{Hest19a} the Coulomb oscillator solutions of the zitter will carry the electron around the nucleus at the distance $\lambda_{e}$ instead of right through it.  
In the ground state of hydrogen the nucleus just sits inside the zitter circle. This is an analog of the Darwin term in wave mechanics. Calculations of the ground state energy are therefore especially sensitive to the model for the nucleus.

At a deeper level, if zitter resonances are characteristic of quantized states, they may play a role in electron-electron interactions. Indeed, it has been suggested \cite{Hest85} that the \textit{Pauli principle} and \textit{Exchange forces} may be explained by zitter resonances. Possibilities for experimental test of those ideas are noted in the many electron theory proposed below.

\subsection{Gravitational binding of the electron}

This Section has presented a well-defined model of the electron as a point charge confined to a superconducting ring. But it raises a natural question: ``What holds the ring together?'' Remarkably, the obvious answer ``Gravity!'' has  strong independent support from General Relativity. 
The main fact is that the celebrated \textit{Kerr-Newman} solution of Einstein's equation involves a charged ring singularity with spin and $g$-factor just like our zitter model of the electron  \cite{Misner70}. That fact has  often been dismissed as a coincidence, but has been given
new currency by arguments for its relevance in elementary particle theory \cite{Burinskii17}.
Without going into details, we note here that our treatment of electron charge density (\ref{5.16}) has been generalized to 
 \textit{Kerr-Newman} by Blinder
himself \cite{Blinder01b}. 
That promises to be a significant step in integrating gravitation with electrodynamics.

\section{From electron to photon}\label
{sec:V}

We have seen how a singularity of the vacuum Dirac equation specified by the Blinder function specifies the seat of the electron's electromagnetic potential. For the sake of possible modification or generalization, let us frame our assumptions in the most general terms. 
We saw in (\ref{4.12}) that the electron wave function $\Psi(x)=\Psi(x-z_e(\tau))$ can be cast in the general form
\begin{equation}
\Psi =R\psi =VUZ^{-1}, \label{5.28a}
\end{equation}
where zitter kinematics is incorporated in the rotor 
\begin{equation}
U=U(\tau)=U_1(\tau)U_2(\tau)
\label{5.28aa}
\end{equation}
given by (\ref{4.12a}), while 
\begin{equation}
\psi = \rho^{1/2}e^{-\i\varphi /2}=Z^{-1},\label{5.28b}
\end{equation}
with $\i =i\bsig_3$, and rotor $ R$ describes the spacetime kinematics of the electron singularity.
Here we propose to interpret $ Z $ as a \textit{complex impedance} of the vacuum, so $ \psi$ is the \textit{vacuum admittance}.  
Considering the role of the Blinder function (\ref{5.16})  relating mass to the vacuum, we propose further to identify $\psi$ with the \textit{Higgs field}, though we do not elaborate that suggestion in this paper.

Let us refer to the multiplicative form (\ref{5.28a}) for a singular solution of the Dirac equation as \textit{vacuum separability}. 
It is noteworthy that the complex admittance $ \psi$ has the form of the Schr\"odinger wave function, which is indeed an approximation to the Dirac wave function \cite{Hest75b,Hest79}. That suggests that Schr\"odinger theory is fundamentally about vacuum singularities. 

Later on we shall consider possibilities for generalizing the ``vacuum impedance.''
To retain essential physical features that we have already identified, we place the following two restrictions on the functional form of $\psi$. First, \textit{vacuum positivity}: 
\begin{equation}
\rho(r)=\psi\tpsi \geq 0, \label{5.31aa}
\end{equation}
with $\rho(r)=0$ only at $r=0$.
Second, for agreement with electrodynamics,
we require that $\rho$ reduce to the Blinder function in the asymptotic region, that is:
\begin{equation}
\rho(r)=e^{-\lambda_{c}/r} \quad\hbox{for} \quad r \gg \lambda_{c} . \label{5.31a}
\end{equation}
These restrictions leave open the possibility
of a more complex functional form for $\psi(r)$ due to short range interactions in the neighborhood of the singularity. 

We note that the Blinder function has an alternative formulation suggesting a general property of vacuum singularities not limited to the electron or even to charged particles. 
We write the electron's Blinder exponent in the form
 \begin{equation}
 \lambda_{c}/r=\frac{e^{2}/\hbar c} 
 {m_{e}c/\hbar v\bdot(x-z_e)}=\frac{\alpha_{e}}{k_{e}\bdot (x-z_e(\tau))}. \label{5.32bb}
 \end{equation}
  This suggests that any particle with kinetic momentum $ k=p/\hbar $ and position $z(\tau)$ will have a Blinder function of the form
 \begin{equation}
 \rho = e^{-\alpha_{e}/k\bdot (x-z(\tau))}, \label{5.33bb}
 \end{equation}
 so the particle is located at $\rho=0,$
and we drop the subscript on $k_e$ to allow $k$ to be a null as well as timelike.
There is no longer a suggestion here that the exponent is the Coulomb potential of a charged particle.
Here the fine structure constant $ \alpha_{e} $ acts  as a kind of general scaling constant for vacuum singularities, so it may play that role even in strong interactions, as argued by MacGregor \cite{MacGregor07}.
Throughout this article we will identify the presence of an electron (or positron) with a zero of its Blinder function.

As explained in \cite{Hest19a}, the positron wave function $\Psi'$ is obtained from electron wave function $\Psi$ by \textit{antiparticle conjugation}: 
\begin{equation}
\Psi\quad\rightarrow \quad \Psi'=\Psi\bsig_1. \label{5.28bc}
\end{equation}
This generates a reflection of the electron's comoving frame (\ref{4.12}) given by:
\begin{equation}
e_{\mu}=e_{\mu}(\tau)\quad\rightarrow\quad
e'_{\mu} =R'\gamma_{\mu}\tR', \label{5.28bd}
 \end{equation}
with $R'=R\bsig_1 =VU\bsig_1$ given by
(\ref{5.28a}).
The net result is a reflection of rest frame observables along the clock vector axis $\e_1=e_1e_0$ given by:
\begin{equation}
\e'_k=-\e_1\e_k\e_1. \label{5.28be}
\end{equation}
This result is needed to relate electrons and positrons to the structure of photons.

\subsection{Photon structure}

Hard on the spectacular successes of Dirac's electron theory, de Broglie applied it to model the photon. De Broglie must have had high hopes for his photon theory, because he took the unusual step of announcing it with fanfare  and immediately translating it into English   \cite{Broglie34}. 
He argued cogently that the \textit{photon must be composed of an electron-positron pair}  described by the Dirac equation.
Nevertheless, despite his deep insight and clever analysis, he was never able to bring his argument to a successful conclusion.
Here we show how to realize de Broglie's proposal with the zitter particle model depicted in Fig. 2 for both electron and positron. 
We simply assume that both electron and positron have
helical paths with a common center, but separated with a spacelike interval less than a Compton wavelength.

As depicted in Fig. 4, in the rest frame of this electron-positron system, the particle motion projects to a circle with zitter radius $\lambda_e$ where electron and positron are located with fixed angular separation $\varphi_0$ and spin angular momentum  $2\s=\hbar\bsig_3$. Lets call this model of photon structure the \textit{photon ring}. 

Actually, since electron and positron have opposite chirality (as indicated in Fig.5a), the timelike component of their paths exactly cancel, leaving only the spacelike segment $e^- e^+$ depicted in Fig 4.

Now we assume that the photon ring with the $e^- e^+$ segment, and the spin and charge stored within it, propagates as a \textit{photon} along a lightlike path with tangent vector $k$,
as depicted in Fig.$5b$.
It is amusing to think of the $e^-$ and the $e^+$ as terminals of a battery that drives a current of constant energy $\hbar \omega$ around the ring.

As the photon moves, it generates an electromagnetic field that can be represented algebraically by a vector potential 
\begin{equation}
 \A= \rho\,\e\, e^{i\boldsymbol{\hat{k}}\varphi}, \label{5.33bca}
\end{equation}
with amplitude $\rho$ and phase  
$\varphi(t) = k\cdot z +\varphi_0 $ 
with $k\cdot z=\omega t-\boldsymbol{k}\cdot \z(t).$
Note that the variables for time $t $ and frequency $\omega$ are set as initial conditions in the instantaneous rest frame of the electron emitting the photon. 
We also assume that the  photon carries energy $\hbar\omega\le 2m_e c^2$ with the limiting value given for pair production (Fig.5$a$).

\begin{figure}
\centering
 \includegraphics[width=0.9
 \linewidth]{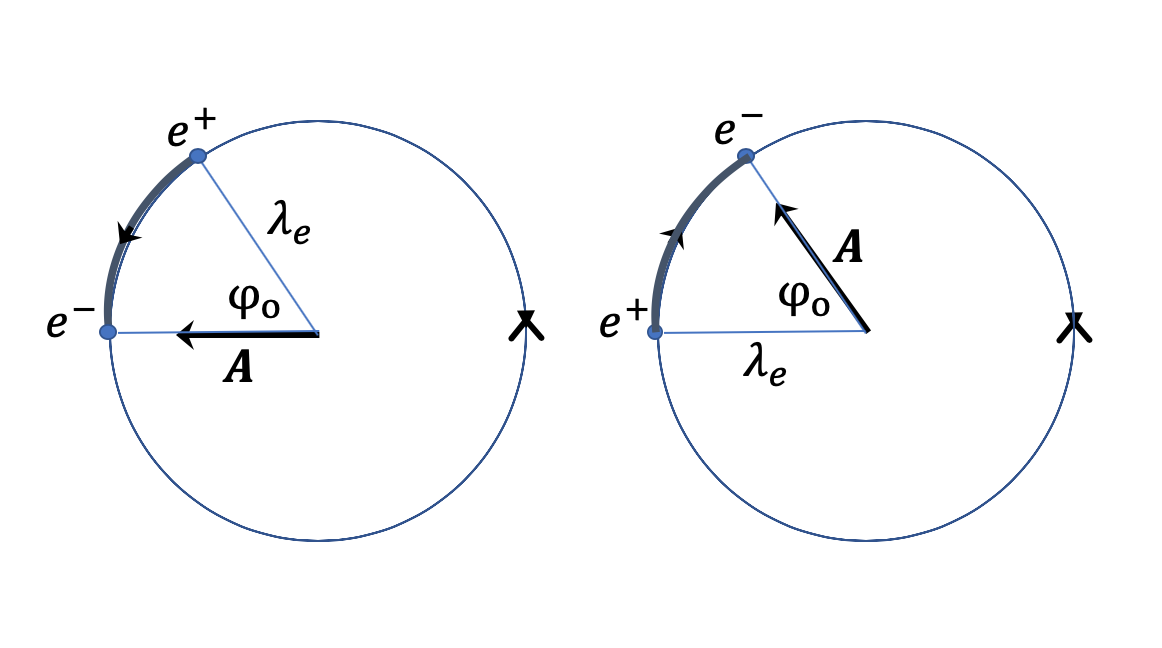}
 \caption{Photon structure: A photon is composed of an electron-positron pair circulating at the speed of light with a fixed angular separation $\varphi_0$, where $0\leq \varphi_0 \leq \pm \pi$ designates left and right circular polarizations.  }
\label{fig:Photon dipole 1}
 \end{figure}

\begin{figure}
\centering
 \includegraphics[width=0.7\linewidth]{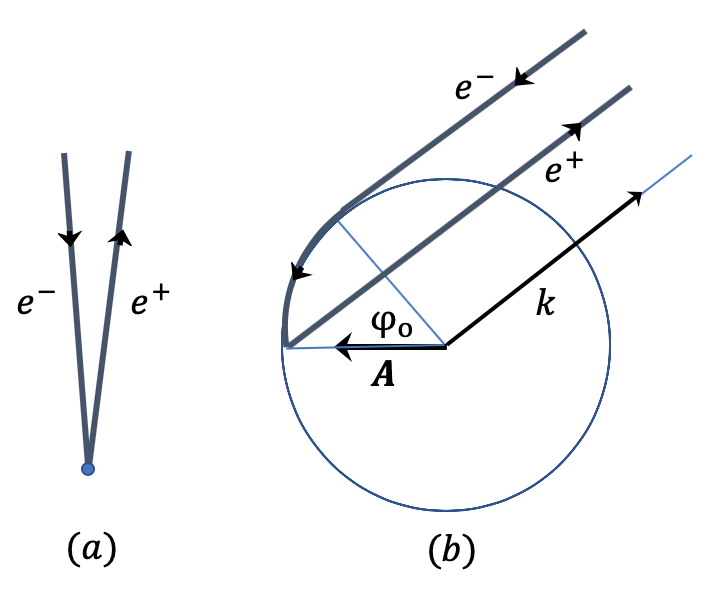}
 \caption{Photon propagation: (a) Feynman diagram for pair creation  (b) Photon propagation is centered along a lightlike path with tangent vector $k$.} 
\label{fig: Photon dipole 2}
 \end{figure}


The photon propagates at the speed of light with momentum $p=\hbar k$, so $p^2=0.$ The photon energy  $\hbar\omega=p\bdot v $ is determined by the proper velocity $v=v(\tau)$ of the source when it is emitted. Therefore, without loss of generality, we can describe photon emission in the instantaneous rest frame of the electron given by  $v=\gamma_0$, so we have the momentum spacetime split $p\gamma_0 =\hbar(\omega+\boldsymbol{k})$.    
Then from the rest frame independence of our photon model we conclude that \textit{ the momentum of the emitted photon $\p=\hbar\boldsymbol{k}$ must be collinear with with electron spin $\s$}. Among other things, this general result accounts for the \textit{headlight effect} in cyclotron radiation.

We complete our picture of the photon by projecting the lightlike path described in Fig.5$b$  into a spacelike path shown in Fig.6. That depicts the photon as a ring that generates a circularly polarized electromagnetic wave with some given frequency as it propagates with velocity $c\boldsymbol{\hat{k}}$.
It is tempting to picture the field generated in each cycle as akin to a smoke ring, so the whole wave train consists of a chain of discrete ``circulating smoke rings'' much like the ``vortex atoms'' proposed by Lord Kelvin in the nineteenth century.
That would be consistent with the experimentally observed countability of photons, but it is a step beyond our present model.

\begin{figure}
\centering
 \includegraphics[width=01.0\linewidth]{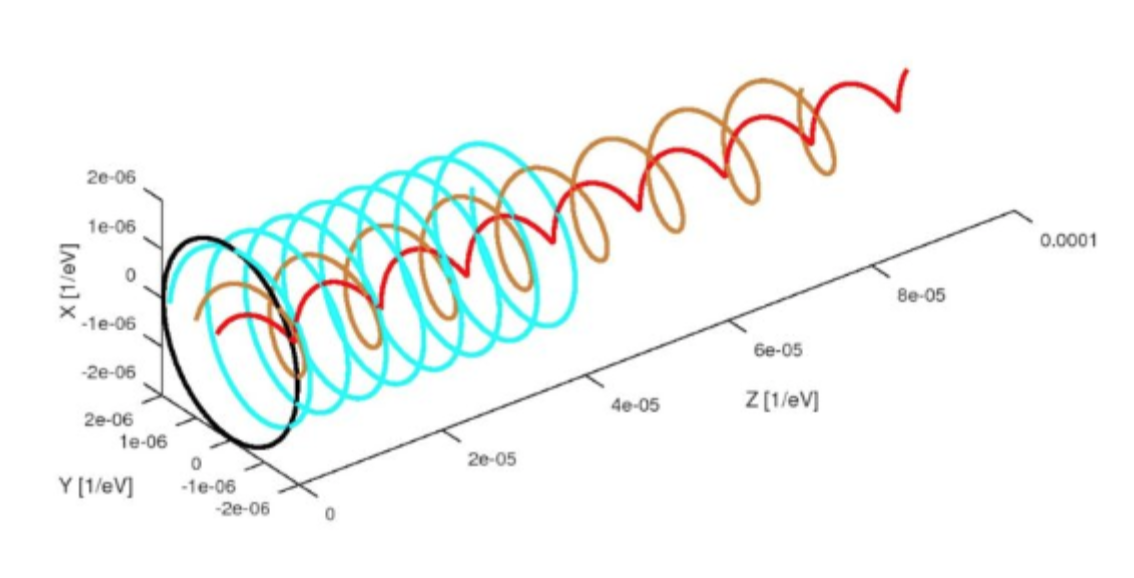}
\caption{ Picture the photon  as a moving ring with angular momentum $\hbar$ that generates an electromagnetic wave with amplitude normalized to its energy. (Figure from  \cite{Vassallo19} with a different but related interpretation)}
\label{fig:Photon radius}
 \end{figure}

The problem remains to square our model of the photon with what is known about electromagnetic radiation discussed in Section IIC. As explained there, the motion of a photon is governed by Maxwell's equation 
\begin{equation}
\partial_0F=-\boldsymbol{\nabla}F ,\label{3.23,repeat}
\end{equation}
with the constraint $F^2 =0.$
For a photon we can write 
$F= \E +i\B=\E(1+\boldsymbol{\hat{k}})$, where the electric field $\E=\partial_0 \A $ is determined by the vector potential given by (\ref{5.33bca}).

The photon field $F=F(\varphi) $  has the same functional form as a plane wave, but its phase function  $\varphi= k\bdot z(\tau)$ is centered on a lightlike curve $z=z(\tau)$. 
Hence, the phase is given by $\varphi=\omega t-\boldsymbol{k}\cdot\z.$
and Maxwell's equation (\ref{3.23})
reduces to the eigenvalue equation
\begin{equation}
  \boldsymbol{k}F=\pm F\omega, \label{5.33bc}
\end{equation}
where the signs correspond to states of left/right circular polarization.
Moreover, we can decompose $F$ into the canonical form
\begin{equation}
F= \E +i\B = \boldsymbol{f}Z 
, \label{3.3ebd1}
\end{equation}
where 
 $Z=Z(\varphi)=\rho e^{i\varphi}$
can be regarded as a complex impedance of the photon singularity, 
and $\boldsymbol{f}$ is a \textit{polarization bivector}
with various forms given by Eqns.
(\ref{2.43k}), (\ref{2.43l}) and (\ref{3.31}).
In particular, we can write
\begin{equation}
\boldsymbol{f}=\hat{\e}\, e^{i\hat{\mathbf{k}}\varphi_0}=\hat{\e}(cos\, \varphi_0 +i\hat{\mathbf{k}}\,sin\, \varphi_0)=\e +i\b, 
 \label{3.3edc22}
\end{equation}
where $\varphi_0 $ is the polarization angle in Fig. 5,
and $\boldsymbol{f}\boldsymbol{f}^{\dagger}=1$.



Like the electron, the photon is a singularity in the electromagnetic vacuum field with  density $\rho=\rho(z)$ plausibly described by (\ref{5.33bb}).
Moreover, it has singularity structure described as a circular ring in Fig.4 and Fig.6.
That gives the photon a size and shape.
One might worry that the photon density (\ref{5.33bb}) could propagate like the electron's Coulomb potential to influence the photon's surroundings in a way that has not been observed. However, it is a general theorem \cite{Doran96} that influence from a null surface, like the boundary of a photon path in the present model, will propagate only along that boundary.
In free space the photon moves in a straight line. However, in a wave guide or optical fiber, the path is shaped by the material walls that modify the parameter $\rho$.

\begin{figure}
\centering
 \includegraphics[width=0.6\linewidth]{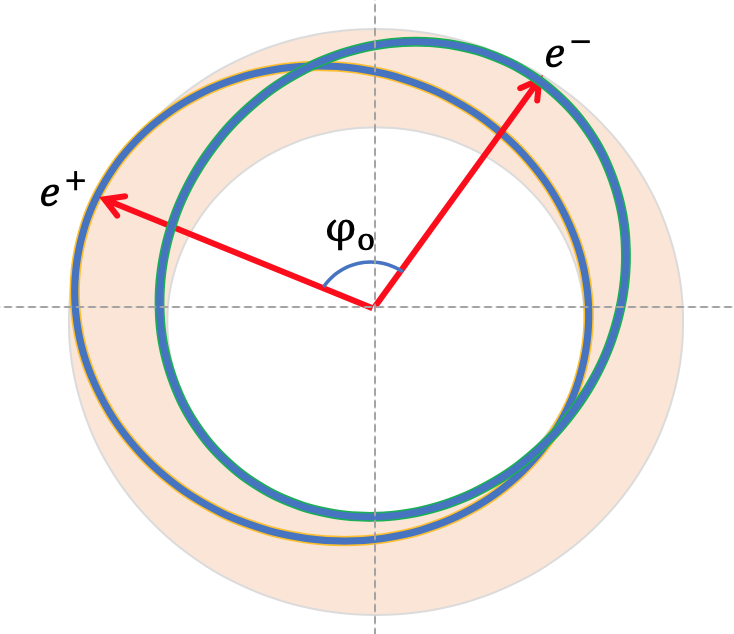}
 \caption{The photon can be modeled as an electron-positron pair located on a toroidal ring (or energy shell)  with a fixed angular separation $\varphi_0$ designating its polarization.  }
\label{fig:Photon dipole 3}
 \end{figure}

To complete our model of photon structure, we note that it can be refined by replacing the circular ring in Fig.4 with a toroidal ring as depicted in Fig.7.  As explained in Section IIIC, the width of the toroidal energy shell is determined by the electron anomalous magnetic moment  $\mu_e$. That may have theoretical implications for binding of the photon, but the more significant point is that it increases the degrees of freedom for the vector potential $\A$ specified in  (\ref{5.33bca}) from $1$ to $2$. Whence, with $\e\bdot\boldsymbol{k}=0$   the polarization vector $\e$ can be 
generated by rotor $U=U_1U_2$ and written
\begin{equation}
\e=U\bsig_1 \tilde{U}
\label{3.3edc24}.
\end{equation}
as specified before by (\ref{7.19}) and (\ref{7.26}).
This provides strong theoretical grounds for predicting the existence of quantized toroidal states for individual photons.
Experimentalists will be proud to announce that they got there first! They have already detected toroidal states in the diffraction of individual photons! 
\cite{Helmerson16}

However, that may be a beginning rather than the end of the story, a presage of a richer landscape of toroidal states in elementary particle theory to be considered in the next Section.

\subsection{Compton Scattering}
 
Compton scattering by a photon on an electron is governed by the conservation law
\begin{equation}
\hbar k +p=\hbar k'+p'.  \label{3.3edc3}
\end{equation}
For an electron at rest we have $p=m_e cv$ with $v=\gamma_0$.
With a simple calculation in \cite{Hest99} this gives us a shift in the photon wavelength:
\begin{equation}
\Delta \lambda=\lambda'-\lambda = \lambda_e(1-\mathbf{\hat{k}\bdot\mathbf{\hat{k}}'}),  \label{3.3edc4}
\end{equation}
where $\lambda_e =h/m_ec$ is the \textit{Compton wavelength}, the diameter of the electron zitter cycle. This is still the most direct and convincing evidence for zitter in electron and photon structure.

It is noteworthy that Compton's equation (\ref{3.3edc4}) seems to be independent of photon polarization. However, the propagation vector $\boldsymbol{k}$ is orthogonal to polarization as expressed by $<\boldsymbol{k}\boldsymbol{f}>= \boldsymbol{k}\bdot\boldsymbol{f}=0$. Hence, the change of polarization in scattering is given by
\begin{equation}
\mathbf{\hat{k}'}\bdot
\mathbf{\hat{k}}=
\boldsymbol{f}'\bdot
\boldsymbol{f}^{\dagger}
=<\boldsymbol{f}'
\boldsymbol{f}^{\dagger}>.  
 \label{3.3edc5}
\end{equation}
This agrees with polarization factors in more elaborate STA treatments of scattering given in 
\cite{Lewis2001}.

 



\section{Vacuum Universality}\label
{sec:VI}

The simplicity and power of modeling the electron as a vacuum singularity strongly suggests that the spacetime vacuum can be regarded as a universal medium for the physical world, so all elementary particles can be regarded as vacuum singularities of various types. 
Success in explaining quantum mechanics and QED for the electron promises strong support for the general thesis that the Dirac equation describes spacetime dynamics of vacuum singularities. Thus we have new prospects for a \textit{unified vacuum field theory} of elementary particles.

With electrons modeled as vacuum singularities, it is natural to consider the topology of more complex vacuum singularities to model the whole zoo of elementary particles, including photons.
A promising possibility is based on the fact that, in a certain sense, the electroweak gauge group is already inherent in the Dirac equation, and a gauge theory version of gravitational interactions is readily included as well. A unified ``gravelectroweak theory'' of that kind has already been  described in \cite{Hest08}. It suffices to summarize its main features here and discuss what Vacuum Dirac theory has to add.

The general idea is that gravity is about deformation of the vacuum due to presence and propagation of singularities described by the Dirac equation.
The implication that all elementary particles and their interactions can be described by variations and excitations of the vacuum impedance promises closure to the search for a Unified Field Theory. 


\subsection{Lepton wave functions}

The electron kinematic rotor  (\ref{5.28aa}) can be expressed in the more explicit form
\begin{equation}
 U(\tau) = e^{-i\bsig_{3}\phi}\, e^{-i\bsig_{2}\theta}. \label{8.13}
 \end{equation}
Now suppose that both angles are harmonically related functions of proper time $\phi(n\tau)$ and $\theta(k\tau)$, where $(n,k)$ is a pair of coprime integers known in knot theory as   \textit{rotation} and \textit{writhe} numbers respectively.
The angle $\varphi$ generates circular zitter in the spin plane, while the angle $\theta$ tilts the plane, so together the two angles can be adjusted to generate a family of closed toroidal curves (or helices) with periods in ratio $k/n$. 
 
Regarding each helix as the path of a point singularity like the electron, we have here a family of singularity types distinguished by quantum numbers $(n,k )$. 
Others have proposed knot theory for classifying elementary particles, though not with such a direct tie 
to the Dirac equation and geometry of the vacuum. Jehle \cite{Jehle71,Jehle75}, for one, proposed a classification based on quantized flux that has much in common with the present approach, but it is weak on connection with field equations.
Finkelstein \cite{Finkelstein07,Finkelstein15} has developed a detailed match of knot topology with structure of the Standard Model. Knot topology can be taken as supplementing the present approach, which is based on differential geometry using STA.

In the following we take standard Electroweak Theory as given, at least in the main. 
The lepton family is composed of three generations: ${(e^-\nu_e) (\mu^-\nu_\mu) (\tau^-\nu_\tau)}$ and their antiparticles.
Our aim is to incorporate it into Maxwell-Dirac theory by identifying the various leptons with their kinematic rotors
in the form given by
(\ref{8.13}).  
Like the electron itself, the kinematic rotor for each lepton has the form
\begin{equation}
U_1(\tau)U_2(\tau) = e^{-i\bsig_{3}\omega_k\tau}\,e^{-i\bsig_{2}\omega_k k\tau}, \label{8.13b}
\end{equation}
where
$\omega_k=m_k c/\hbar$ is a constant frequency specified by each leptonic mass $m_k$, and each generation $\{e, \mu, \tau\}$ is indexed by its \textit{writhe} number $k=1, 2, 3$. Why the number of generations is limited to three is unknown, though it may be due to a mass constraint. A natural geometric explanation for the quantum number $k$ has been already noted in Consa's model for toroidal zitter.

A fundamental feature of Electroweak Theory is that each charged lepton is paired with a unique neutrino companion.
As noted in \cite{Hest85}, this fits with the fact there are only two possibilities for the path of a zbw center: timelike or lightlike. 
It conforms nicely with Barut's proposal \cite{Barut86} that the electron carries the neutrino within itself along with its electromagnetic field.

An attractive possibility is that the energy in the electrons anomalous magnetic moment might be carried away by a neutrino much like $beta$-decay.
Indeed, if circular zitter is an intrinsic geometric property of the electron, as suggested by (\ref{7.15}),
then it may well be decoupled from charge to produce a neutral particle.
There seems to be no alternative account for the neutrino, because we have used up all degrees of freedom in the wave function. Moreover, this leads by analogy to testable experimental predictions about muon and tau neutrinos.


This completes our discussion of kinematic rotors for modeling leptons. But a few words are in order to fit it into Barut's account of elementary particles \cite{Barut80,Barut86}.

In the Standard Model of elementary particles, 
the three generations of leptons stand in obvious analogy to the three generations of quarks. That suggests that they must be related in some way. Since the leptons are observable particles while quarks are not,
the simplest possibility is that all hadrons are composed of leptons.
The quarks then represent symmetries in the lepton composition rather than actual particles. 

As noted by Barut, to explain nucleons as composed of leptons, we must explain \textit{isospin} and \textit{strangeness} quantum numbers in terms of leptons. 
To do that, Barut observed that the number of $\mu^\pm$ mesons in hadrons is exactly equal to the ``\textit{strangeness}'' quantum number in hadrons. Evidently ``\textit{strange}'' hadrons decay into ordinary hadrons if the $\mu$ inside the hadron  decays. In strong interactions $\mu$ is stable, hence strangeness is conserved.

Barut goes on to generalize the electron's Coulomb potential to include   magnetic interaction that dominates in ranges less than a Compton wavelength. This remains a work in progress to describe   fundamental interactions of the electron in electromagnetic terms. Presumably, it can be incorporated into a generalization of the Blinder function.

\subsection{Gravelectroweak Gauge Theory}

A natural extension of the Dirac equation to include weak interactions rests on the unique fact that the electroweak gauge group SU(2)$\otimes$U(1) is the maximal group of gauge transformations $\Psi\rightarrow \Psi'=\Psi U$ that leave the velocity observable invariant:
\begin{equation}
\rho u=\Psi \gamma_{0}\widetilde{\Psi} =\Psi' \gamma_{0}\widetilde{\Psi'} ,\label{8.10}
\end{equation}
This gives the gauge group geometric significance as the invariance group of the Dirac current, thereby insuring a spacetime path for the zitter center. 
To incorporate both gravitational and electroweak interactions in vacuum Dirac theory, we require invariance under the group
\begin{equation}
\Psi \quad \rightarrow \quad \Psi'=L\Psi U, \label{8.11}
\end{equation}
where $L\widetilde{L}=1$. The corresponding gauge invariant derivative is
\begin{equation}
D_{\mu}{\Psi}=(\partial_{\mu}+\half\omega_{\mu})\Psi- \Psi iW_{\mu}, \label{8.12}
\end{equation}
where the geometric ``connexion'' $\omega_{\mu}$ expresses gravitational interaction and the $W_{\mu}$  express electroweak interactions. See \cite{Hest08} for an extensive account of the details.

Applied to modeling the electron, the gauge derivative (\ref{8.12})
leads \cite{Hest08} to a generalization of the \textit{Kerr-Neuman} solution of Einstein's equation with a Blinder function yet to be determined.
This promises to be a pathway uniting gravity with electromagnetism.

\subsection{Stable Composite Particles}

The upshot of Electroweak Theory is that all leptons are  just different states of the electron.
This suggests that all elementary particles and resonances can be built out of electrons.
That possibility had been studied at some depth by Asim Barut \cite{Barut80,Barut86} until his unfortunate sudden death.
Now the insights into electroweak gauge theory outlined in the preceding Section open up new possibilities.

As Barut observed, aside from the electron and neutrino, the only \textit{stable elementary particles} are the photon and proton. 
It follows that the whole spectrum of elementary particles (and antiparticles) can be generated by leptons if the photon and proton can. Let us consider how the zitter model of electron structure may make that possible.

\subsubsection{Proton structure} 

Given the model of the photon as an $(e^+e^-)$ pair, 
we have an obvious extension to a proton model $(e^+e^-e^+)$ simply by inserting a positron centered at the CM with its circulating charge located on the torus of the photon. Presumably, in the proton bound state the three circulating charges are distributed on the torus in some symmetrical way that generates the proton's magnetic moment. The similarity of this toroidal coupling with the three quark model of the proton is noteworthy, and would seem not to require gluons for binding.

\section{Many Particle Theory}\label
{sec:VII}

As a minimal generalization of \textit{Vacuum Dirac Theory} to a many particle theory,  we consider a system of $N$ electrons regarded as particle singularities in the vacuum with zitter velocities $u_{k}$ and CM spacetime paths 
$ z_{k}=z_{k}(\tau_{k}) $ with proper velocities 
$ v_{k}=\dot{z}_{k} $. We suppose their motions are determined by a spinor wave function for the vacuum $ \Psi=\Psi (x,z_{1},z_{2},. . . ,z_{N}) $. The vacuum density is then given by 
\begin{equation}
\Psi\widetilde{\Psi}=\rho=\prod^{N}_{k=1}\rho_{k}=\check{\rho}_{k}\rho_{k},\label{8.1}
\end{equation}
where $ \check{\rho}_{k} $ designate the product with the  $k$th factor omitted, and, as before, 
\begin{equation}
\rho_{k}=e^{-\alpha_{k}},\label{8.2}
\end{equation}
where $\alpha_{k}=\alpha (r_{k})$ is a Blinder potential or its generalization
with retarded position $ r_{k}=(x-z_{k})\bdot v_{k} $.

As the basic equation for energymomentum density in the vacuum, we consider a straightforward generalization of the Gordon current in Born-Dirac theory, namely
\begin{equation}
\rho P=\frac{e}{c}\sum^{N}_{k=1}A_{k}\check{\rho}_k=\rho \sum^{N}_{k=1}p_{k},    \label{8.3}
\end{equation}
where, as before,
\begin{equation}
\frac{e}{c}A_{k}=\rho_{k}p_{k}=m_{e}c\rho_{k} u_{k}-\square\bdot(\rho_{k} S_{k})\label{8.4}
\end{equation}
is the electromagnetic \textit{vector potential} (Gordon current) of the $k$th particle modeled with or without zitter, and 
\begin{equation}
\rho P_{\mu}=\hbar\langle (\partial_{\mu} \Psi)\i\widetilde{\Psi}\rangle    \label{8.5}
\end{equation}
defines components $ P_{\mu}=\gamma_{\mu}\bdot P $ of the \textit{canonical momentum.} 

Since equation (\ref{8.3}) is the core synthesis of Maxwell's electrodynamics with Dirac's electron theory, consolidating what we have discussed already and providing a platform for extensions to follow, let me christen it with the name \textit{Maxwell-Dirac Equation}.

Generalization to include other fermions has been discussed already. Restricting our attention to electrons for the moment, we note that equation (\ref{8.3}) has obvious implications for the Helium atom, where it will treat both electrons on equal footing and imply correlations similar to the ``exchange interaction.'' Carrying out the calculations would provide a stringent test of (\ref{8.3}) with implications for the Pauli principle.

Equation (\ref{8.3}) also meets Carver Mead's objective for a many electron quantum state determined entirely by vector potentials of all particles in the system \cite{Mead97,Mead2000}.  
It goes beyond Mead in anchoring the electron state in a Dirac wave function,  in principle including the contribution of positive charges in the lattice of a superconductor. Note that  
$ P $ in (\ref{8.3}) can be regarded as kind of ``\textit{superpotential}" for the entire system. It follows, then, that 
\begin{equation}
F\equiv \square\wedge P= \sum^{N}_{k=1}\square\wedge p_{k}    \label{8.5a}
\end{equation}
can be regarded as the total electromagnetic field for the entire system. 

Inside a superconductor we have $ F= \E+i\B =0$ (Meissner effect). Hence, as we have seen before, Stokes Theorem implies that
for \textit{any closed curve} in the region
\begin{equation}
  \oint P\bdot dx=0. \label{8.5b}
\end{equation}
And, as in the single particle case, we get a many particle quantization condition
\begin{equation}
  \int_{0}^{T_{n}}P_{0}\,dt =\oint \mathbf{P}\bdot d\x=(n+\half)h, \label{8.5c}
\end{equation}
 with integer $ n $. This agrees with Mead's formulation of phase and flux quantization in a superconductor  \cite{Mead97,Mead2000}.
The relevance of this argument to the \textit{Aharonhov--Bohm effect} is also worth noting \cite{Post95}.


Specific application to superconductors is beyond the purview of this exploratory discussion. However, before dropping the subject, it is worth noting that the present model satisfies the additivity of electron phases $ \varphi_{k}=\varphi_{k}(x-z_{k}) $ that is essential for superconductivity. 
That can be made manifest by writing the wave function in the form
\begin{equation}
\Psi=R\prod^{N}_{k=1}\Psi_{k}\Lambda,
\label{8.6}
\end{equation}
where $ R\tR=1,\; \Lambda^2=0 $ and
\begin{equation}
\Psi_{k}=e^{-\alpha_{k}-\i\varphi_{k}}.
\label{8.7}
\end{equation}
Evidently, the \textit{Pauli principle} can be incorporated in symmetries of the wave function in the usual way, and that would identify it as a property of vacuum singularities! The symmetries need not apply to all particle variables, but only to particles whose motions are resonant in some sense, as in the quantized atomic states discussed earlier.

\subsection{Particle Diffraction}

Maxwell-Dirac theory has  unique implications  for the problem of electron diffraction, indeed, for particle diffraction in general. We point them out here without delving into detailed calculations or experimental tests.

The first and most important point is that, according to (\ref{8.1}), the density $\rho =\rho(x)$ of a single electron factors into a product
\begin{equation}
\rho=\prod^{N}_{k=1}\rho_{k}
=\check{\rho}_{e}\rho_{e},\label{8.8a}
\end{equation}
where $\rho_{e}(x)$ is the Blinder function of Maxwell-Dirac theory, and 
$\check{\rho}_{e}(x) = \check{\rho}_{e}(x,x_1,\dots,x_{N-1} )$ is the  density  expressed with Blinder functions of all other particles with influence. Since the Blinder function satisfies $0\le|\rho_k|\le 1$, we also have $0\le|\rho(x)|\le 1$. So there are no normalization issues, and sufficiently distant particles automatically have insignificant influence.  

One consequence is that the  ``Quantum  potential'' in  the Pilot Wave   guidance law  must have a causal source in 
matter composing the diffraction slits.
To get the ``acausal density'' of Pilot Wave theory, the matter coordinates must be integrated out with some sort of average $\langle\rho\rangle$. That leaves the possibility open for  fluctuations in path density, for example, from heating material in the slits. 

We still have the problem of identifying  a plausible mechanism for momentum exchange between each diffracted particle and the slits, a causal link which is missing from all accounts of diffraction by standard wave mechanics or by Pilot Wave theory.
Note that momentum transfer is observable for each scattered particle, whereas the diffraction pattern conserves momentum only as a statistical average.
Evidently the only way to account for this fact is by reducing diffraction to quantized momentum exchange between each particle and slit.
To that end, \cite{Mobley18} provides a detailed analysis of optical diffraction patterns explained by photon momentum exchange.

Duane was the first to offer a quantitative explanation for electron diffraction as quantized momentum exchange \cite{Duane23}. A more general argument using standard quantum mechanics has been worked out by Van Vliet \cite{Vliet63,Vliet10}. These explanations suffer from the same disease as Old Quantum Mechanics in failing to account for the density distribution in the diffraction pattern. However, we now have the possibility of curing that disease with relativistic Pilot Wave theory.
We only need to explain how the momentum exchange is incorporated into the Pilot Wave guidance law.

Now, presuming vanishing  electric and magnetic fields outside the diffraction slits as before, we have $\square \wedge A=0$, so locally, at least, $A$ is a gradient. Assuming the same for $P$, we have a gauge invariant \textit{phase
 gradient }
\begin{equation}
\square \Phi =P-\frac{e}{c}A.\label{8.8c}
\end{equation}
This provides a promising mechanism for quantized momentum transfer in diffraction. For we know that quantized states in QM are determined by boundary conditions on the phase. Successful calculation of diffraction patterns along these lines would provide strong evidence 
for the following claim: the vacuum surrounding electromagnetically inert matter is permeated by a vector potential with vanishing curl. Remarkably, the same mechanism would explain the extended Aharonov-Bohm (AB) effect \cite{Batelaan10}.
Evidently, then, the causal agents for diffraction and the AB effect are one and the same: a universal vector potential
permeating the vacuum (or, \textit{Aether}, if you will) of all spacetime, much as proposed by Dirac \cite{Dirac51b}.

Considering  the similarity of electron and photon diffraction patterns, we should expect the same mechanism to explain both, especially if photons are composed of electron-positron pairs as proposed in  Section V.  Indeed, the evolution of path density for the electron is determined by the Dirac equation, which gives
\begin{equation}
\square^2 \Phi=-m_e c
\dot{z}\cdot \square \ln \rho.
\label{8.8ff}
\end{equation}
For a photon with propagation vector $k$, the analog is
\begin{equation}
k\cdot \square \ln \rho =
\square^2 \Phi/\hbar,
\label{8.8f}
\end{equation}
where, of course, $\rho$ is the path density for photons, just as it is for electrons. 
Accordingly, we conclude that diffraction is ``caused'' by the vacuum surrounding material objects. In other words, \textit{diffraction is refraction by the vacuum}!

We have seen that the Blinder form for the vacuum density $\rho=\rho(x)$, which was originally introduced to generalize the Coulomb potential, is actually determined by the momentum at each vacuum singularity independent of any charge. Evidently it applies to photons as well as electrons and protons, so it should   be regarded as a universal property of the vacuum. This suggests association with a gravitational field. That possibility is best approached by a gauge theory as proposed in Section VB.

Strictly speaking, the density (impedance) of the vacuum should be incorporated into any vector potential by writing 
$ \mathcal{A}=\rho A$, with a new notation to distinguish it from the usual vector potential, whether or not it is the gradient of a scalar field. The Aether can then be regarded as a conserved fluid (with $\square\cdot\mathcal{A}=0)$
flowing through spacetime with particle  singularities (electron, photon or whatever) in the density swept along.
This picture has a beautiful macroscopic analog describing diffraction of a macro particle in a classical fluid
\cite{Bush15}.


\label{F.8}


\subsection{Vacuum Topology}

Derivation of equations of motion for singularities from gravitational field equations has been an important theoretical objective since Einstein, Infeld and Hoffman attacked it \cite{Einstein39}. We have seen something like that for Dirac theory. Conversely, if we develop a rich theory of singularities along the lines suggested here, that might require modification of the field equations.
Gauge theory may then be regarded as a means for coordinating singularities with equations of motion.

``Still keeping one principal object in view, to preserve their symmetrical shape!''


\setlength{\textheight}{682pt}

\bibliographystyle{hunsrt}
\bibliography{zbw}

\end{document}